\documentclass[prd, preprint, nofootinbib, showpacs]{revtex4}

\usepackage{epsf,epsfig,subfigure,graphicx,amsmath,amssymb}
\usepackage{amsfonts}
\usepackage{latexsym}
\usepackage{color}

\newcommand{\dis}[1]{\begin{equation}\begin{split}#1\end{split}\end{equation}}
\newcommand{\be}{\begin{equation}}
\newcommand{\ee}{\end{equation}}
\def\bea{\begin{eqnarray}}
\def\eea{\end{eqnarray}}

\newcommand{\eq}[1]{Eq.~(\ref{#1})}
\newcommand{\eqs}[2]{Eqs.~(\ref{#1}) - (\ref{#2})  }

\newcommand{\bfrac}[2]{{\left(\frac{#1}{#2} \right)  }}\newcommand{\VEV}[1]{\langle #1 \rangle}

\newcommand{\Mp}{M_P}
\newcommand{\calP} {{\cal P}}
\newcommand{\tilder} {{\tilde r}}

\newcommand{\fnl}{f_{\rm NL}}
\newcommand{\gnl}{g_{\rm NL}}
\newcommand{\taunl}{\tau_{\rm NL}}

\newcommand{\Qs}{Q_\sigma\sigma_*}
\newcommand{\Qss}{Q_{\sigma\sigma}\sigma_*^2}
\newcommand{\Qsss}{Q_{\sigma\sigma\sigma}\sigma_*^3}

\begin{document}

\title{
 Modulated reheating by curvaton 
}

\author{Ki-Young Choi}
 \affiliation{
 Asia Pacific Center for Theoretical Physics, Pohang, Gyeongbuk 790-784, Republic of Korea and \\
 Department of Physics, POSTECH, Pohang, Gyeongbuk 790-784, Republic of Korea}

\author{Osamu Seto}
 \affiliation{
 Department of Life Science and Technology,
 Hokkai-Gakuen University,
 Sapporo 062-8605, Japan
}

%

\begin{abstract}
%
There might be a light scalar field during inflation which is not responsible
 for the accelerating inflationary expansion.
Then, its quantum fluctuation is stretched during inflation.
This scalar field could be a curvaton, if it decays at a late time.
In addition,
 if the inflaton decay rate depends on the light scalar field expectation value by interactions between them, 
 density perturbations could be generated by the quantum fluctuation of the light field when the inflaton decays. 
This is modulated reheating mechanism.
We study curvature perturbation in models
 where a light scalar field does not only play a role of curvaton
 but also induce modulated reheating at the inflaton decay. 
We calculate the non-linearity parameters
  as well as the scalar spectral index and the tensor-to-scalar ratio. 
We find that there is a parameter region where
 non-linearity parameters are also significantly enhanced
 by the cancellation between the modulated effect and the curvaton contribution.
For the simple quadratic potential model of both inflaton and curvaton, 
 both tensor-to-scalar ratio and nonlinearity parameters could be simultaneously large.
%
\end{abstract}


\preprint{APCTP-Pre2012-006, HGU-CAP-015} 

\vspace*{3cm}
\maketitle


\section{Introduction}

Cosmic inflation solves various problems
 in the standard Big Bang cosmology~\cite{Inflation}
 and simultaneously
 provides the seed of large scale structure in our Universe from
 the quantum fluctuation of a light scalar field, e.g., inflaton field $\phi$~\cite{InflationFluctuation}.

A single field inflation model predicts
 the density perturbation which is  nearly scale-invariant and almost Gaussian.
In other words, the scalar spectral index $n_s$ is close to unity and
 a non-linearity parameter $f_{\rm NL}$ is  
 much less than unity~\cite{Maldacena:2002vr}.
This is consistent with the current limit on the local type non-linearity 
 parameter $\fnl$ from the 
 Wilkinson Microwave Anisotropy Probe (WMAP) seven-year data,
 $-10<\fnl<74$ at the 95\% confidence level~\cite{Komatsu:2010fb}. 

However, besides canonical single field slow-roll inflation models,
 there are many possible mechanisms to generate density perturbation.
By WMAP data, the scalar spectral index $n_s$ has been measured with a good accuracy,
 while the non-linearity parameters have been just weakly constrained as above.  
The sensitivity of the Planck satellite~\cite{Planck} to measure non-Gaussianity
 is as good as to probe $f_{\rm NL}$ of ${\cal O}(1)$.
The non-Gaussianity could be an important observable
 to discriminate between various mechanisms of density perturbation generation.
 
For example, multi-field inflation models can show large non-Gaussianity with special conditions
 during inflation~\cite{Choi:2007su,Byrnes:2008wi,Byrnes:2008zy,Kyae:2009jm,Elliston:2011dr,Choi:2011me}, 
 at the end of inflation~\cite{Lyth:2005qk,Alabidi:2006wa,Sasaki:2008uc,Naruko:2008sq,Choi:2012he},
 preheating~\cite{preheating}, or deep in the radiation dominated era~\cite{Byrnes:2010em}.
The last case includes the ``curvaton''
 scenario~\cite{Mollerach:1989hu,Linde:1996gt,CurvatonLW,CurvatonMT,CurvatonES}.
A light scalar field, curvaton, has too little potential energy
 to drive inflationary expansion during inflation.
At a later time when a curvaton decays, 
 the isocurvature perturbation of the curvaton field becomes adiabatic
 or mixed with that from the inflaton field.
If the curvaton energy density is subdominant at its decay time,
 the large non-Gaussianity is generated in general~\cite{CurvatonNG}.
Since an inflaton also generates density perturbations,
 inflaton and curvaton contributions to density perturbation could be comparable.
This mixed inflaton-curvaton scenario has been also
 studied intensively~\cite{MixedIC}.

The quantum fluctuation of a subdominant light scalar field during inflation
 can modulate the efficiency of reheating by the inflaton decay~\cite{Dvali:2003em,Kofman:2003nx}. 
This makes the reheating a spatially inhomogeneous process. 
The quasi-scale invariant perturbations of this field,
 which are isocurvature modes during inflation,
 may be converted into the primordial curvature perturbation
 during this process. 
Large non-Gaussianity also can be induced from modulated
 reheating~\cite{Zaldarriaga:2003my,Bauer:2005cd,Suyama:2007bg,Ichikawa:2008ne}. 
For a review of modulated reheating after inflation, see for example Ref.~\cite{Bassett:2005xm}.
Again, in general,
 the density fluctuation can be generated by both inflaton and modulated reheating.
Mixed inflaton-modulated reheating scenario also has been investigated~\cite{Ichikawa:2008ne}.

Here one may realize that a curvaton is a light scalar field and
 thus naturally can play the role of a scalar field which modulates the reheating by the inflaton decay.
The inflaton field may have a coupling with a curvaton
 if that is small enough not to disturb dynamics of both an inflaton and a curvaton.
Nevertheless this interaction modifies the decay rate of the inflaton field.  
Since the decay rate of inflaton becomes a function of the local value of a curvaton
 field $\sigma ({\bf x})$,
 this gives rise to a perturbation in the decay rate of the inflaton field
 and thus in the reheating temperature which is responsible for the density
 perturbation after reheating. 
So far few attention has been paid to couplings between the inflaton and the curvaton~\cite{Suyama:2010uj},
 compared to self-interaction of curvatons~\cite{Enqvist:2010dt,CurvatonSelf}.
 In this study,
 we incorporate the modulated reheating effects in the curvaton scenario,
 taking the perturbation generated from the inflaton field also into account.

The paper is organized as follows.
After describing the model and its dynamics in section~\ref{dynamics},
 we consider the curvature perturbation and calculate
 the power spectrum, the tensor-to-scalar ratio and non-linearity parameters in section~\ref{perturbation}. 
In section~\ref{models} we work on simple models and
 show how the modulated reheating effect by curvaton affects the density perturbation
 and parameter space of models can be constrained. 
We summarize our results in section~\ref{conclusion}.

\section{Dynamics}
\label{dynamics}

Inflation is driven by the potential energy of the inflaton field, $\phi$.
We assume that during inflation interaction terms of inflaton with other fields are negligible.
However after inflation, the inflaton starts oscillating around the minimum and finally,
 via the interaction terms, decays into the standard model (SM) particles,  which makes the hot thermal plasma
 in the standard Big Bang cosmology.

One of the relevant terms for the decay of inflaton field including curvaton field $\sigma$
 could be given by
\begin{equation}
 {\cal L}_{\rm int} = \lambda |\Phi |^2 \phi \sigma ,
\label{PhiSigmaInteraction}
\end{equation}
 where $\Phi$ is another light scalar field such as the SM Higgs field.
Then, in the classical background of curvaton field, 
 Eq.~(\ref{PhiSigmaInteraction}) induces the decay of inflaton
 into two Higgs scalars, with the curvaton expectation value dependent (CD) decay width 
\begin{equation}
 \Gamma_{\phi}^{CD}(\sigma) = \frac{1}{8 \pi m_{\phi} } \lambda^2 \sigma^2 ,
\end{equation}
 with $m_{\phi}$ being the inflaton mass at the minimum.
Together with the other curvaton independent interactions such as
\begin{equation}
 {\cal L} = {\cal O}_{\rm SM} \frac{\phi}{M_P} ,
\label{NRInteraction}
\end{equation}
 where ${\cal O}_{\rm SM}$ denotes a SM operator,
 induce the curvaton independent (CI) decay width of the inflaton, $\Gamma_{\phi}^{CI} $.
The total decay rate of the inflaton is given by
\begin{equation}
 \Gamma_{\phi} (\sigma)=  \Gamma_{\phi}^{CI} + \Gamma_{\phi}^{CD}(\sigma) .
\label{Deay:total}
\end{equation}
For the very light curvaton field, $ \Gamma_{\phi} > m_{\sigma}$, with $m_\sigma$ being curvaton mass,
 the curvaton starts to oscillate in the radiation-dominated epoch well after the inflaton decays,
 when the Hubble parameter becomes as small as $m_{\sigma}$.
After the reheating by the inflaton $\phi$ is completed,
 the energy density of the radiation from inflaton decay decreases as
\begin{equation}
 \rho_r = 3 \Mp^2\Gamma_{\phi}^2 \left(\frac{a_{\Gamma_{\phi}}}{a}\right)^4, \label{rhor}
\end{equation}
where $a_{\Gamma_{\phi}}$ is a scale factor when inflaton decays, i.e. $H=\Gamma_\phi$. 
The energy density of the curvaton after the onset of the oscillations decreases  as
\begin{equation}
 \rho_{\sigma} = \frac{1}{2} m_{\sigma}^2 \sigma_*^2 \left(\frac{a_{m_{\sigma}}}{a}\right)^3, 
 \label{rhosigma}
\end{equation}
 with $\sigma_*$ being the expectation value of curvaton during inflation and $a_{m_{\sigma}}$
 is a scale factor when the curvaton start oscillation at $H=m_\sigma$.
The curvaton decays at a later time and we call its decay rate $\Gamma_{\sigma}$.
Whether the Universe is curvaton dominated or radiation dominated
 at the moment of curvaton decay depends on the size of $\Gamma_{\sigma}$ and $\sigma_*$.

\section{Primordial curvature perturbation}
\label{perturbation}

We consider that inflaton $\phi$ and curvaton $\sigma$ fields are relevant
 to the density perturbation in the early Universe. 
Their vacuum fluctuations are promoted to a classical perturbation around the time of horizon exit. 
During inflation the field trajectory is dominated by the inflaton field and
 thus the inflaton perturbation becomes adiabatic and that of the curvaton contributes
 to the isocurvature mode. 

\subsection{Modulated reheating from an interaction with a curvaton}

Reheating of the Universe is attained from the decay of the inflaton field $\phi$.
Since the inflaton decay is modulated by the curvaton field $\sigma$, the curvature perturbation
 of the radiation produced from the decay of inflaton has two origins.
One comes from  the inflaton field itself in the standard picture of
the generation of fluctuations. 
The other comes from the light scalar field (curvaton) $\sigma$ during the reheating process
 due to the interaction between the inflaton and the curvaton field.
Thus, as in the inflaton-modulated reheating mixed scenario,
 it is written~\cite{Zaldarriaga:2003my, Ichikawa:2008ne} as
\begin{eqnarray}
\zeta_r
 &=& \frac{1}{\Mp^2}\frac{V}{V_\phi}\delta\phi_* + \frac{1}{2\Mp^2} \left( 1- \frac{VV_{\phi\phi}}{V_\phi^2} \right)\delta\phi_*^2
 + \frac{1}{6 M_P^2}\left( -\frac{V_{\phi\phi}}{V_{\phi}} -\frac{V V_{\phi\phi\phi}}{V_{\phi}^2}
 +2\frac{V V_{\phi\phi}^2}{V_{\phi}^3}  \right) \delta\phi_*^3  \nonumber \\
 && +Q_\sigma\delta\sigma_* + \frac12Q_{\sigma\sigma}\delta\sigma_*^2
 + \frac{1}{6}Q_{\sigma\sigma\sigma}\delta\sigma_*^3 +\cdots ,
\end{eqnarray}
 where $Q$ is a function of $\Gamma_\phi(\sigma)/H_c$ calculated  at a time $t_c$ which  is after several oscillations of the inflaton
 but well before the time of decay of inflaton. 
A quantity with $*$ is evaluated  when the corresponding scale crosses the Hubble horizon during inflation.

During inflation the energy density of $\sigma$ field is negligible and
 the inflation is driven by the inflaton field $\phi$ alone.
The slow-roll parameters during inflation are defined by
\dis{
 \epsilon\equiv \frac{\Mp^2}{2}\bfrac{V_\phi}{V}^2,\qquad 
 \eta \equiv \Mp^2 \frac{V_{\phi\phi}}{V},\qquad 
 \xi^2 \equiv \Mp^4 \frac{V_\phi V_{\phi\phi\phi}}{V^2}.
\label{slow-roll}
}
Using Eq.~(\ref{slow-roll}),  the curvature perturbation of radiation $\zeta_r$ is expressed as
\dis{
\zeta_r= \zeta_{r1} + \frac{1}{2}\zeta_{r2} + \frac{1}{6}\zeta_{r3} +\cdots ,
\label{zetar}
}
with
\begin{eqnarray}
\zeta_{r1} &=& \frac{1}{\Mp\sqrt{2\epsilon_*}} \delta\phi_* +Q_\sigma \delta\sigma_* ,  \\
\zeta_{r2} &=& \frac{1}{\Mp^2} \left( 1- \frac{\eta_*}{2\epsilon_*} \right)\delta\phi_*^2 + Q_{\sigma\sigma} \delta\sigma_*^2 ,  \\
\zeta_{r3} &=& \frac{1}{ M_P^3 \sqrt{2\epsilon_*}}
 \left( -\eta - \frac{\xi^2}{2 \epsilon} + \frac{\eta^2}{\epsilon}  \right)\delta\phi_*^3
 + Q_{\sigma\sigma\sigma} \delta\sigma_*^3 .
 \label{zetar123}
\end{eqnarray}

\subsection{After curvaton decay}

After inflaton decay,
 the energy density of radiation decreases, however the curvaton energy density stays the same for a while
 and starts to decrease when the mass of curvaton becomes larger than the Hubble expansion.
Since the energy density of oscillating curvaton decreases slower than that 
 of radiation, the curvaton becomes important well after the decay of inflaton.
When the decay rate of curvaton $\Gamma_\sigma$ becomes comparable to the Hubble expansion rate,
 the curvaton $\sigma$ decays quickly to radiation.

After the decay of the curvaton, the remnant radiation is a mixture from the inflaton and curvaton
decay products
 with different density perturbations. 
In this inflaton-curvaton mixed scenario,
 the curvature perturbation after the curvaton decay can be expressed as analytically
 with instant decay approximation by~\cite{Bartolo:2003jx,Lyth:2005fi,Sasaki:2006kq},
\begin{equation}
 \zeta = \zeta_1+\frac{1}{2}\zeta_2+\frac{1}{6}\zeta_3+\ldots,
\label{zeta}
\end{equation} 
where
\dis{
\zeta_1=&(1-R)\zeta_{r1} +R\zeta_{\sigma 1},\\
\zeta_2=&(1-R)\zeta_{r2} +R\zeta_{\sigma 2}
+R(1-R)(3+R)\left(\zeta_{r1}-\zeta_{\sigma 1}\right)^2,\\
\zeta_3=&(1-R)\zeta_{r3} +R\zeta_{\sigma 3}
+3R(1-R)(3+R)\left(\zeta_{r1}-\zeta_{\sigma
  1}\right)\left(\zeta_{r2}-\zeta_{\sigma 2}\right)\\
+& R(1-R)(3+R)(-3+4 R +3 R^2)\left(\zeta_{r1}-\zeta_{\sigma 1}\right)^3,
}
 with
\begin{equation}
 R \equiv \left.\frac{3 \rho_{\sigma} }{4 \rho_r + 3 \rho_{\sigma}}\right|_{H=\Gamma_{\sigma}} . \label{R}
\end{equation}
 Here $\zeta_{r}$ is given in \eq{zetar} and $\zeta_{\sigma}$ is the curvature perturbation
 from the curvaton field. 
$R$  parametrizes the dominance of the curvaton energy density when it decays.

For the quadratic potential of curvaton, $\zeta_\sigma$ is given by~\cite{Sasaki:2006kq}
\begin{eqnarray}
\zeta_\sigma &=& \zeta_{\sigma 1} +\frac{1}{2}\zeta_{\sigma 2} +\frac{1}{6}\zeta_{\sigma 3} + \cdots 
\nonumber \\
&=& \frac23 \frac{\delta \sigma_*}{\sigma_*}
- \frac13 \bfrac{\delta\sigma_*}{\sigma_*}^2 
  +\frac29\bfrac{\delta\sigma_*}{\sigma_*}^3 +\cdots.
  \label{zetasigma}
\end{eqnarray}
Then from \eq{zeta} with \eqs{zetar123}{zetasigma} we obtain
\begin{eqnarray}
\zeta_1 &=& \frac{1-R}{\Mp\sqrt{2\epsilon_*}} \delta\phi_* +\left( (1-R)Q_\sigma + \frac{2R}{3\sigma_*} \right) \delta\sigma_*, \label{zeta1}\\
\zeta_2 &=& (1-R)\left\{ \frac{1}{\Mp^2} \left( 1- \frac{\eta_*}{2\epsilon_*} \right)\delta\phi_*^2
 + Q_{\sigma\sigma} \delta\sigma_*^2\right\}
 -\frac{2R}{3}\bfrac{\delta\sigma_*}{\sigma_*}^2 \nonumber \\
 && +R(1-R)(3+R)\left(  \frac{\delta\phi_*  }{\Mp\sqrt{2\epsilon_*}} 
 + Q_{\sigma} \delta\sigma_*  - \frac23 \frac{\delta \sigma_*}{\sigma_*}\right)^2,\label{zeta2}\\
\zeta_3 &=& \frac{1-R}{\Mp^3\sqrt{2\epsilon_*}}  \left( -\eta - \frac{\xi^2}{2 \epsilon} + \frac{\eta^2}{\epsilon}  \right)\delta\phi_*^3+(1-R)Q_{\sigma\sigma\sigma} \delta\sigma_*^3
+\frac{4R}{3}\bfrac{\delta\sigma_*}{\sigma_*}^3  \nonumber \\
&&
+3R(1-R)(3+R)\left(  \frac{\delta\phi_*  }{\Mp\sqrt{2\epsilon_*}} 
 + Q_{\sigma} \delta\sigma_*  - \frac23 \frac{\delta \sigma_*}{\sigma_*}\right)  \nonumber \\
&&\times
 \left\{\frac{1}{\Mp^2} \left( 1- \frac{\eta_*}{2\epsilon_*} \right)\delta\phi_*^2 + Q_{\sigma\sigma} \delta\sigma_*^2+\frac23\bfrac{\delta\sigma_*}{\sigma_*}^2\right\} \nonumber \\
&& + R(1-R)(3+R)(-3+4 R +3 R^2)\left(  \frac{\delta\phi_*  }{\Mp\sqrt{2\epsilon_*}} 
 + Q_{\sigma} \delta\sigma_*  - \frac23 \frac{\delta \sigma_*}{\sigma_*}\right)^3.
 \label{zeta3}
\end{eqnarray}

\subsection{The power spectrum}

The power spectrum $\calP_\zeta$ of the curvature perturbation is defined by
\dis{
\VEV{\zeta(k_1)\zeta(k_2)} =(2\pi)^3  \delta ({\bf k}_{1}+{\bf k}_{2} )
 \frac{2\pi^2}{k^3}\calP_\zeta  (k_1) , \label{DefPzeta}
 }
and the perturbations of the fields at the horizon exit satisfy
\dis{
\VEV{\delta \phi_*(k_1)\delta\phi_*(k_2)  } &= (2\pi)^3  \delta ({\bf k}_{1}+{\bf k}_{2} )
 \frac{2\pi^2}{k^3}\calP_{\delta\phi_*}  (k_1),\\
 \VEV{\delta \sigma_*(k_1)\delta\sigma_*(k_2)  } &= (2\pi)^3  \delta ({\bf k}_{1}+{\bf k}_{2} )
 \frac{2\pi^2}{k^3}\calP_{\delta\sigma_*}  (k_1),\\
 \VEV{\delta \phi_*(k_1)\delta\sigma_*(k_2)  } &= 0  , \label{twopoint}
}
with 
\dis{
\calP_{\delta\phi_*}  (k)= \calP_{\delta\sigma_*}  (k)= \bfrac{H_*}{2\pi}^2, \label{Pexit}
}
which is determined at around horizon exit.

Using \eq{zeta1} and \eqs{DefPzeta}{Pexit}, the power spectrum of the curvature perturbation is given by
\dis{
 {\cal P}_{\zeta} =& \frac{(1-R)^2}{2\Mp^2\epsilon_*}\calP_{\delta\phi_*} +\left[ (1-R)Q_\sigma + \frac{2R}{3\sigma_*} \right]^2\calP_{\delta\sigma_*} \\
 =&\frac{1}{2\Mp^2\epsilon_*} \bfrac{H_*}{2\pi}^2 (1-R)^2(1+\tilder),\label{Pzeta}
}
 with
\dis{
\tilder \equiv \frac{2\Mp^2 \epsilon_*}{9\sigma_*^2(1-R)^2} \left[ 3Q_\sigma\sigma_*(1-R) + 2R \right]^2.
}
Here $0\leq R \leq 1$ defined in \eq{R} parametrizes the contribution of the curvaton. 
For $R=1$, which means that the curvaton dominates the background energy density when it decays,
the power spectrum is given by 
\dis{
 {\cal P}_{\zeta,0} \equiv \bfrac{H_*}{3\pi\sigma_*}^2.
}

Therefore the contribution to the power spectrum from the curvaton $\sigma$ dynamics scales $R^2$,
 while those from both the inflaton $\phi$ and
 the modulated reheating effect through the $\sigma$ field by inflaton $\phi$ decay scale $(1-R)^2$.
As one can see, in the limit of $R \rightarrow 0 $,
 the usual curvaton contribution disappears and it corresponds to the inflaton-modulated mixed scenario. 
The opposite limit with $R \rightarrow 1 $ corresponds to
 the pure curvaton scenario.
Whereas the $R$ parametrizes the $\sigma$ contribution as the curvaton compared with the other two,
 the $\sigma$ contribution via the modulated inflaton decay
 is parametrized by $Q_{\sigma}$.

The parameter $\tilder$ compares the contribution to the Power spectrum
 from the $\sigma$ field to that from inflaton $\phi$.
In the limit of $\tilder \rightarrow 0$, the Power spectrum comes solely from the inflaton,
 while in the limit of $\tilder \gg 1$ the $\sigma$ field contributes dominantly
 through the modulated effect
 and/or the curvaton effect with  the inflaton effect suppressed.

\begin{figure}[!t]
  \begin{center}
  \begin{tabular}{c}
   \includegraphics[width=0.5\textwidth]{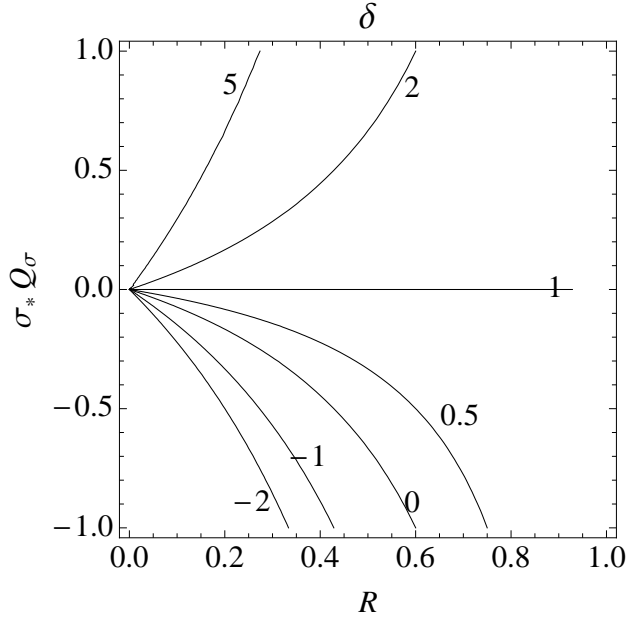}  \end{tabular}
  \end{center}
  \caption{ The contour plot of $\delta$ defined in \eq{delta}. The cancellation happens along the line of $\delta=0$. }
  \label{fig:delta}
\end{figure}

One should notice that because the latter two contributions come from the single same source $\sigma$,
  there is a cross term of  two.  
This cancellation between the modulation effect and the curvaton effect
 makes non-trivial features in the Power spectrum. 
Both contributions may cancel each other when $3 Q_{\sigma} \sigma_* \simeq -2R/(1-R)$
 and the inflaton contribution dominates. 
 To quantify the amount of the cancellation, we define $\delta$ as
 \dis{
 \delta \equiv 1+  \frac{3\Qs (1-R)}{2R}. \label{delta}
 }
This $\delta$ is a measure of fine tuning of the cancellation and becomes $\delta =0$ for the exact cancellation.  The contour plot of $\delta$ is shown in figure~\ref{fig:delta}.
 With this, $\tilder$ can be written as
 \dis{
 \tilder =  \frac{8\Mp^2 \epsilon_*R^2}{9\sigma_*^2(1-R)^2} \delta^2. \label{tilderdelta}
 }

From \eq{Pzeta}, 
 ignoring the negligible contribution from the curvature of the curveton potential $V_{\sigma\sigma}$,
 the scalar spectral index $n_s$ is given by
\begin{eqnarray}
 n_s -1 \equiv \frac{d {\cal P}_{\zeta}}{d \ln k} = -2\epsilon_* +\frac{-4\epsilon_*+2\eta_*}{1+\tilde{r}} ,
 \label{index}
\end{eqnarray}
 which we normalize to be $0.97$ through in our analysis. 
The tensor-to-scalar ratio is given by
\begin{eqnarray}
r_T  \equiv \frac{{\cal P}_T}{{\cal P}_{\zeta}} = \frac{16 \epsilon}{(1-R)^2 (1+\tilde{r})},
\end{eqnarray}
where we used ${\cal P}_T=8(H_*/2\pi)^2$.
As you can see here, for small $\tilder$,
 the observational limit $r_T< 0.36$ constrains the value of $R$ to be
$R< 1-\sqrt{16\epsilon_*/0.36}\simeq 0.53$.

\begin{figure}[!t]
  \begin{center}
  \begin{tabular}{cc}
   \includegraphics[width=0.5\textwidth]{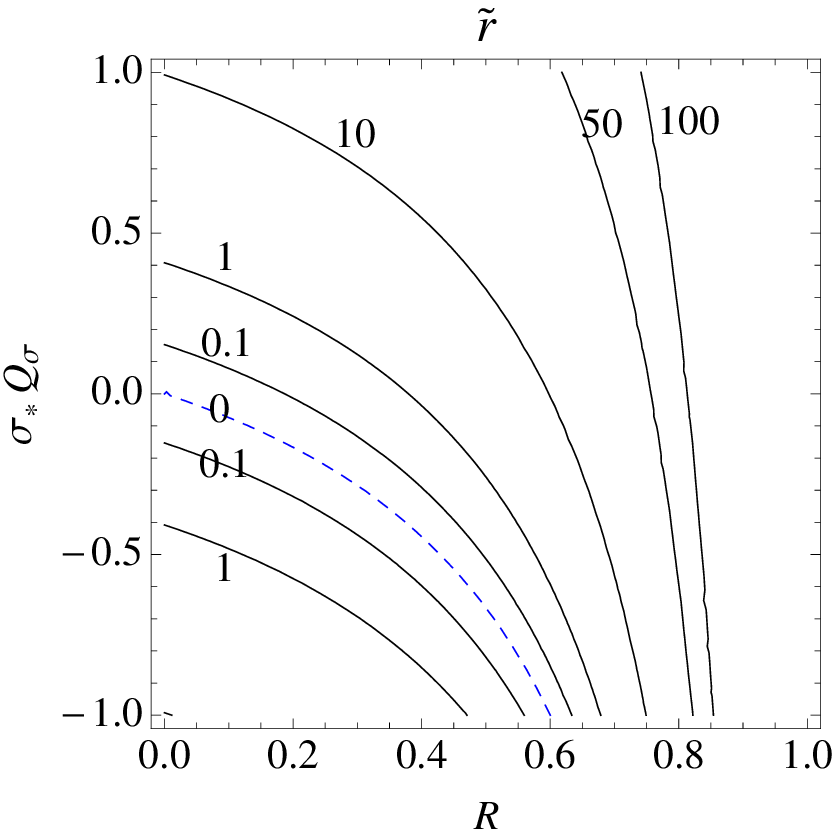}
  & 
\includegraphics[width=0.5\textwidth]{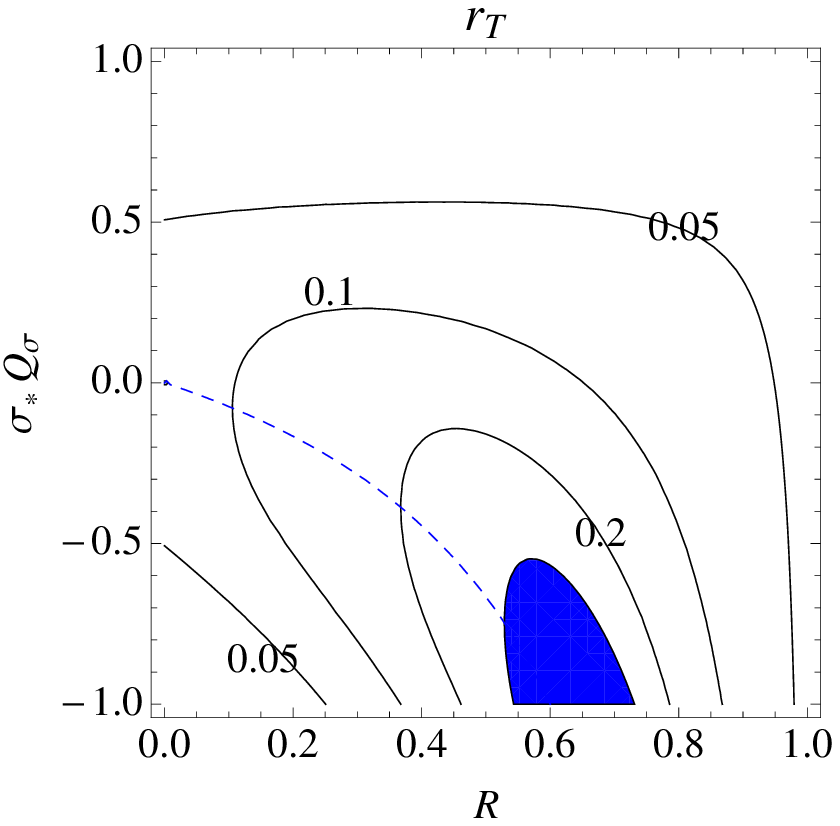} 
  \end{tabular}
  \end{center}
  \caption{[{\bf Left window]} : The contour plot of $\tilder$ for  Case B,  $\sigma_*=0.05\Mp$.  
For the other cases the magnitudes are scaled by $\sigma_*^2$.  Along the blue dashed line $\tilder=\delta=0$ .
{\bf [Right window]} : The tensor-to-scalar ratio $r_T$ for Case B.  The blue shaded region is ruled out using the constraint $r_T<0.36$. $\delta=0$ along the blue dashed line.}
  \label{fig:tilder2}
\end{figure}

In figure~\ref{fig:tilder2},
 we show the contour plot of $\tilder$ for $\sigma_*=0.05\, \Mp$ (which we will later call Case B)
 with $\epsilon_*\simeq 0.005$ and $\eta_*=0$. 
There is a cancellation between curvaton effects and modulated effects
 around the dashed line (blue) which connect $(R,\Qs)=(0.6,-1)$ and $(0,0)$, where $\tilder$ vanishes. 
In this small $\tilder$ limit the inflaton contribution dominates the power spectrum. 
In the opposite region with a large $\tilder$, the $\sigma$ field dominates the Power spectrum.
For different values of $\sigma_*$, the magnitude scales as $\sigma_*^{-2}$,
 since $\epsilon_*$ does not change much.

\begin{figure}[!t]
  \begin{center}
  \begin{tabular}{c c}
   \includegraphics[width=0.5\textwidth]{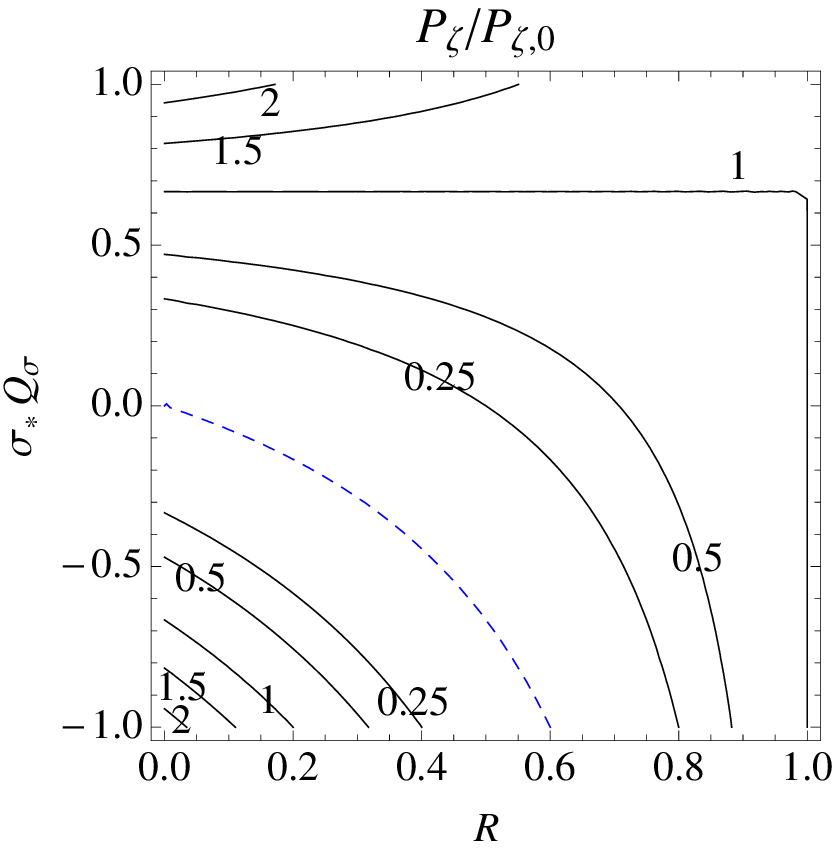}
    & 
  \includegraphics[width=0.5\textwidth]{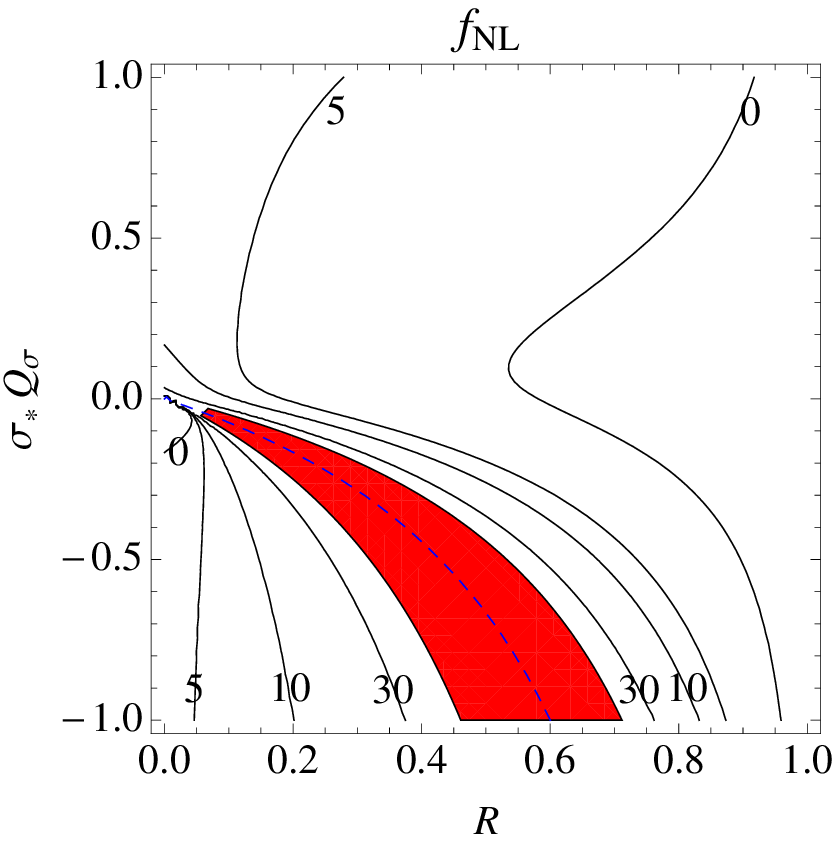} \\
   \includegraphics[width=0.5\textwidth]{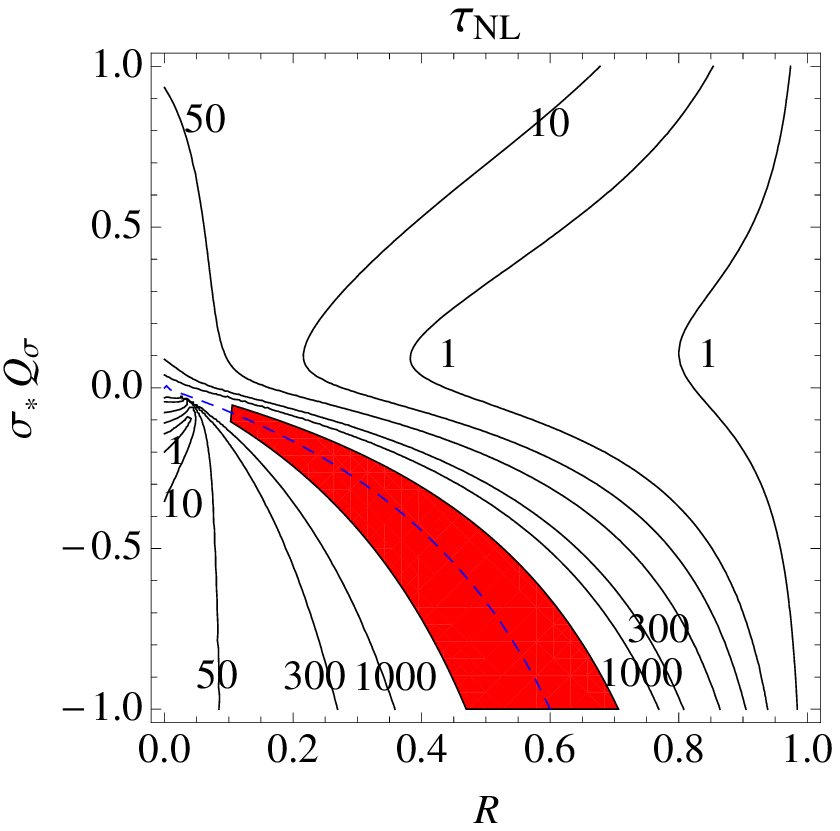}
    & 
  \includegraphics[width=0.5\textwidth]{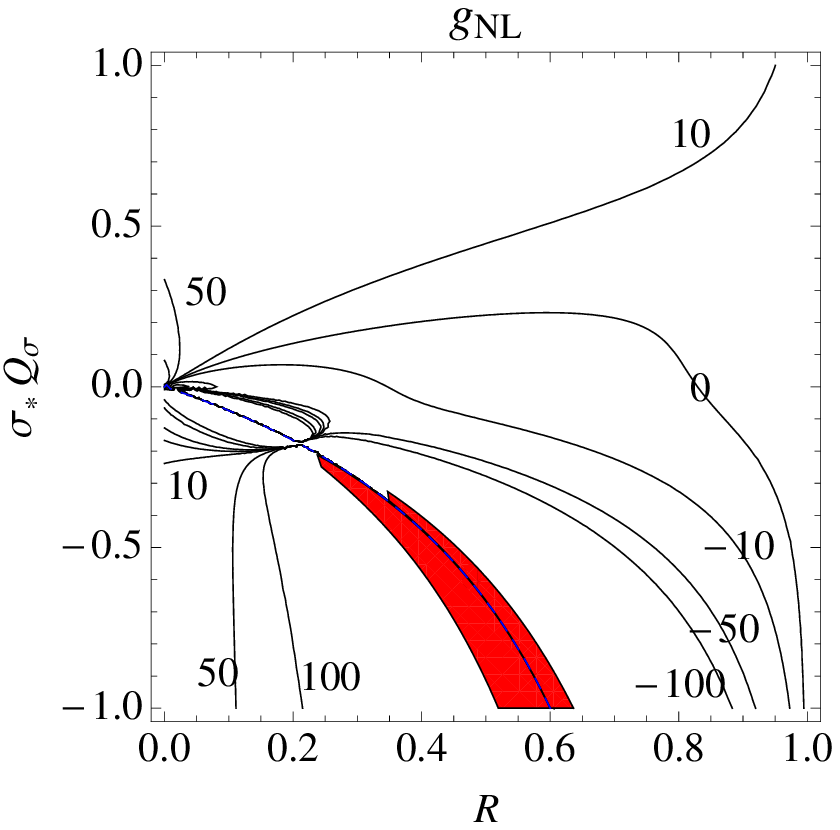}  
  \end{tabular}
  \end{center}
  \caption{The contour plots on $R- Q_{\sigma}\sigma_*$ plane, 
  of the power spectrum ${\cal P}_{\zeta}$  normalized by the value at $R=1$ (upper-left), $\fnl$ (upper-right), $\taunl$ (lower-left)
  and $\gnl$ (lower-right) for Case A, $\sigma_*=10^{-3}\, \Mp$. We put $\eta_*=\xi_*=0$.
  The red shaded region corresponds to too large non-Gaussianity  to be consistent with the observation, $-10<\fnl<73$, $\taunl < 10^4$, and $|\gnl| < 10^5$. Along the blue dashed line $\tilder=0$ the cancellation happens.}
  \label{fig:Case1}
  \end{figure}
\begin{figure}[!t]
  \begin{center}
  \begin{tabular}{c c}
   \includegraphics[width=0.5\textwidth]{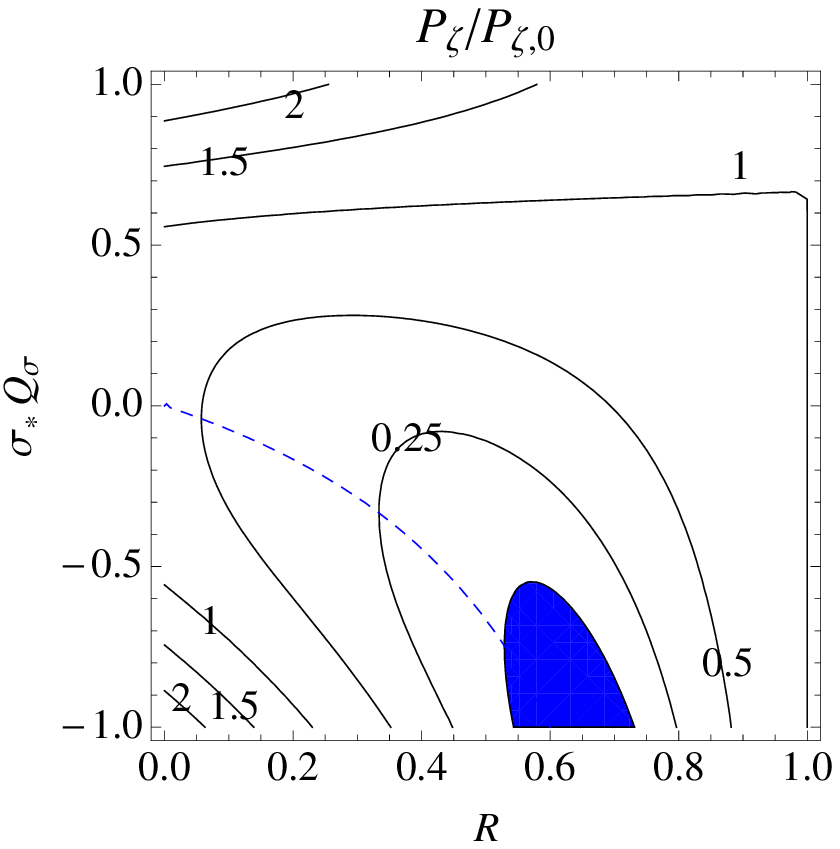}
    & 
  \includegraphics[width=0.5\textwidth]{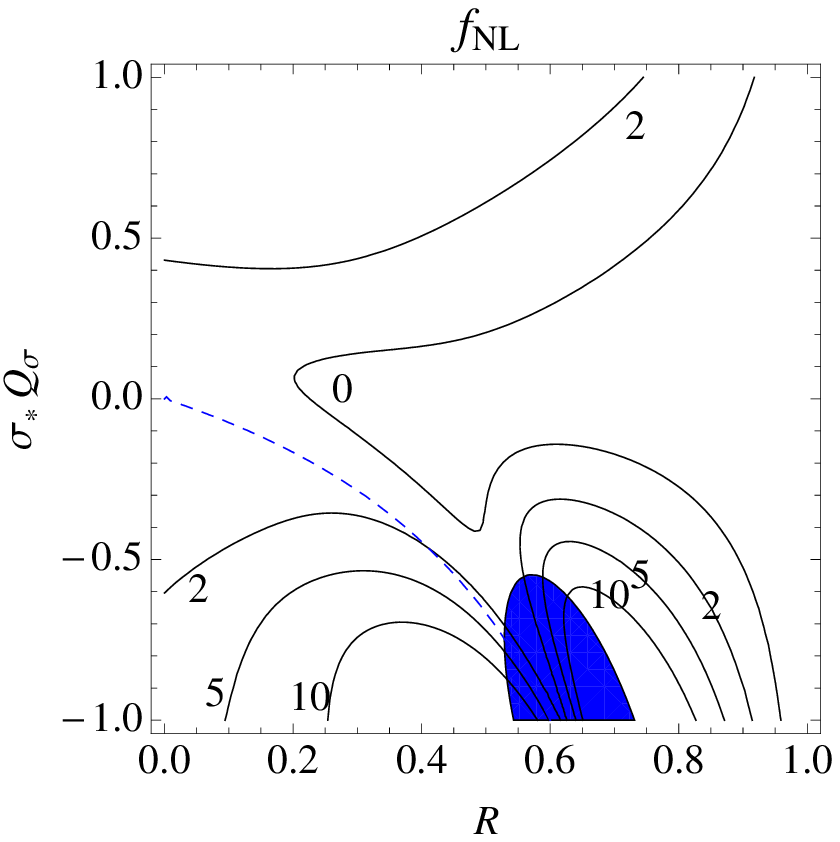} \\
   \includegraphics[width=0.5\textwidth]{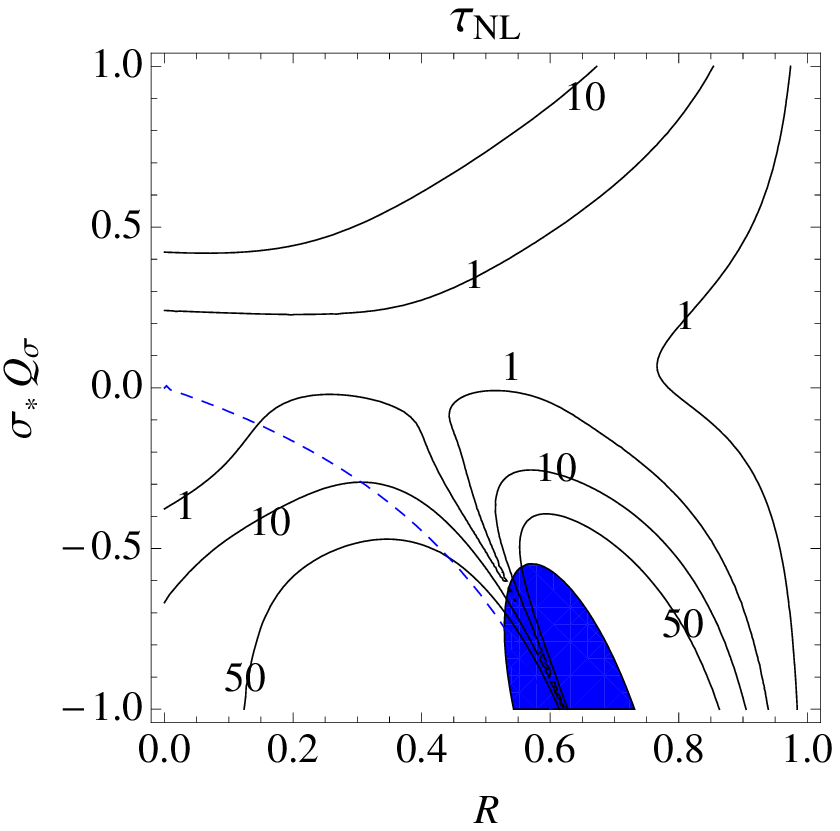}
    & 
  \includegraphics[width=0.5\textwidth]{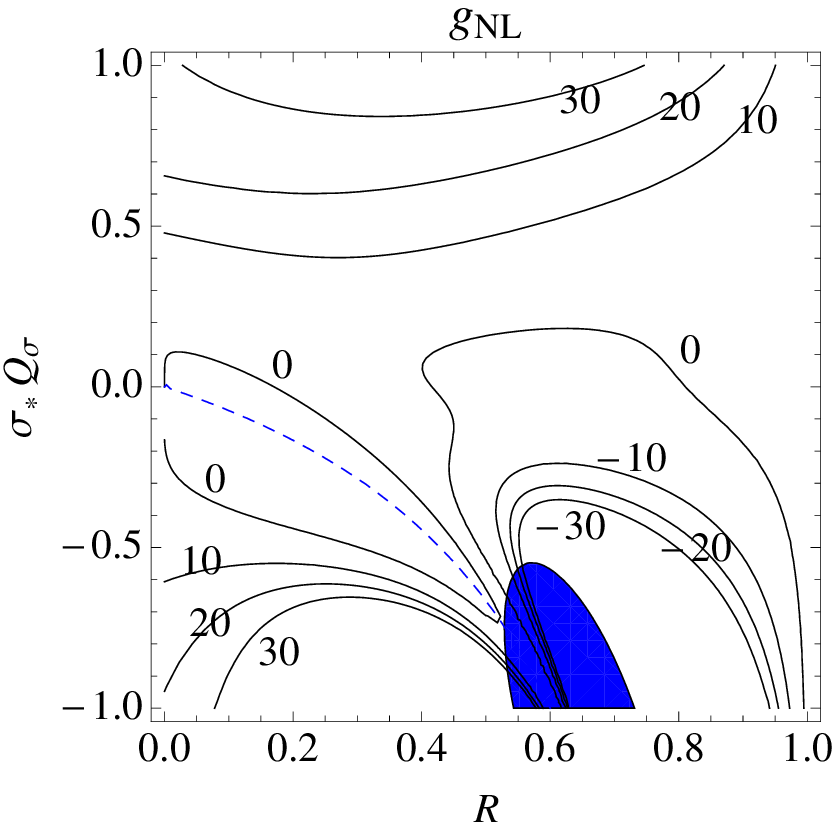}  
  \end{tabular}
  \end{center}
  \caption{The contour plots on $R- Q_{\sigma}\sigma_*$ plane, 
  of the power spectrum ${\cal P}_{\zeta}$  normalized by the value at $R=1$ (upper-left), $\fnl$ (upper-right), $\taunl$ (lower-left)
  and $\gnl$ (lower-right) for Case B, $\sigma_*=0.05\, \Mp$. We put $\eta_*=\xi_*=0$.
  The blue shaded region corresponds to too large tensor-to-scalar ratio to be consistent with the observation, $r_T<0.36$. Along the blue dashed line $\delta=0$.}
  \label{fig:Case2}
\end{figure}
\begin{figure}[!t]
  \begin{center}
  \begin{tabular}{c c}
   \includegraphics[width=0.5\textwidth]{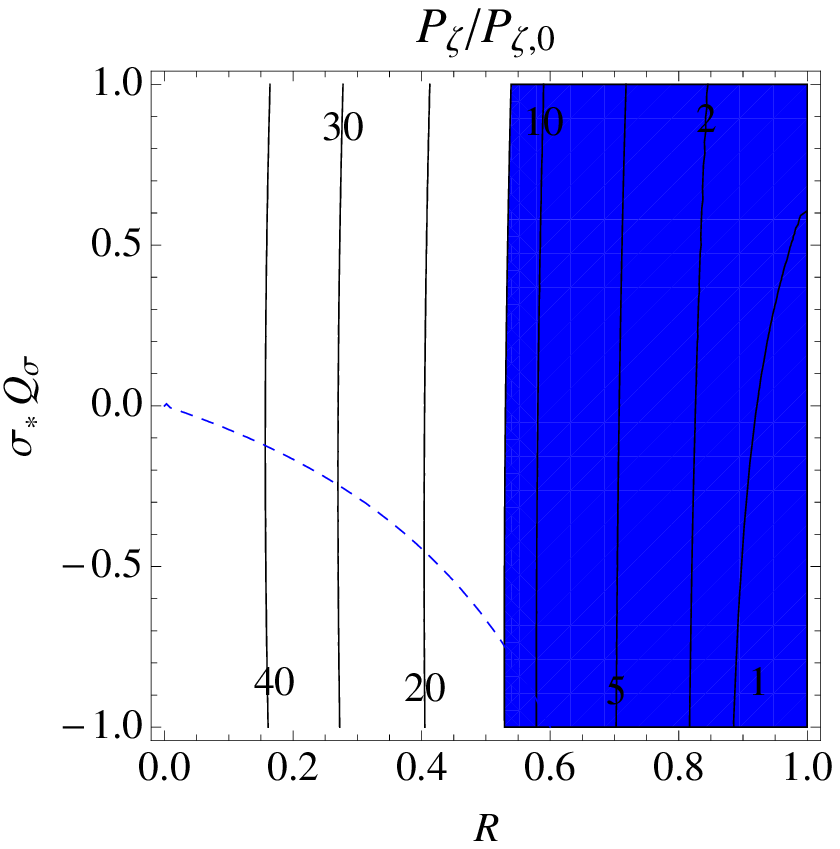}
    & 
  \includegraphics[width=0.5\textwidth]{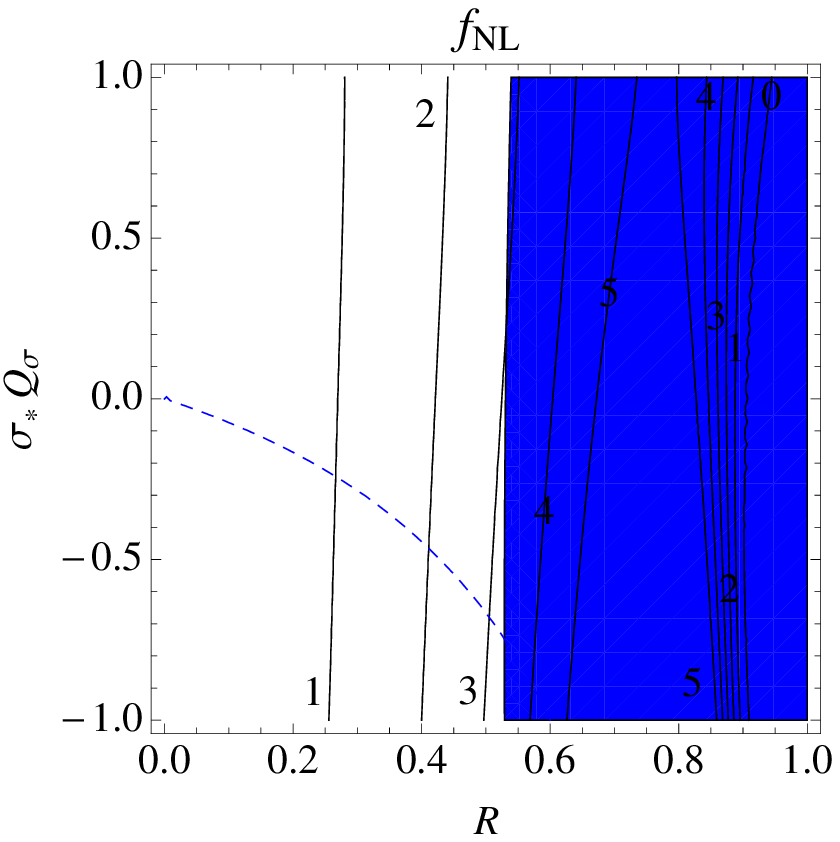} \\
   \includegraphics[width=0.5\textwidth]{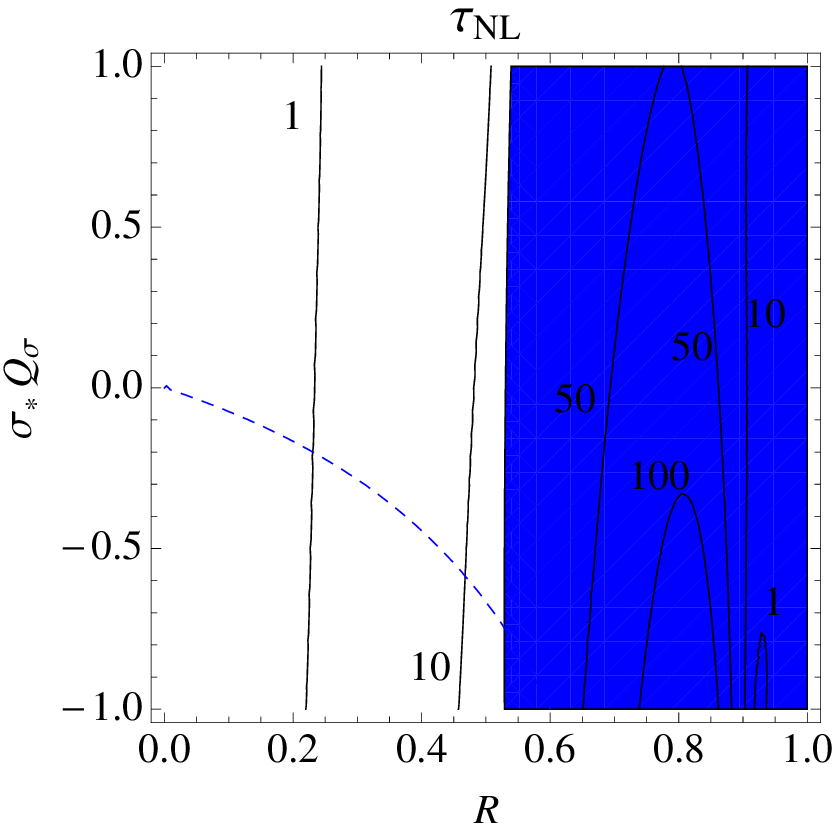}
    & 
  \includegraphics[width=0.5\textwidth]{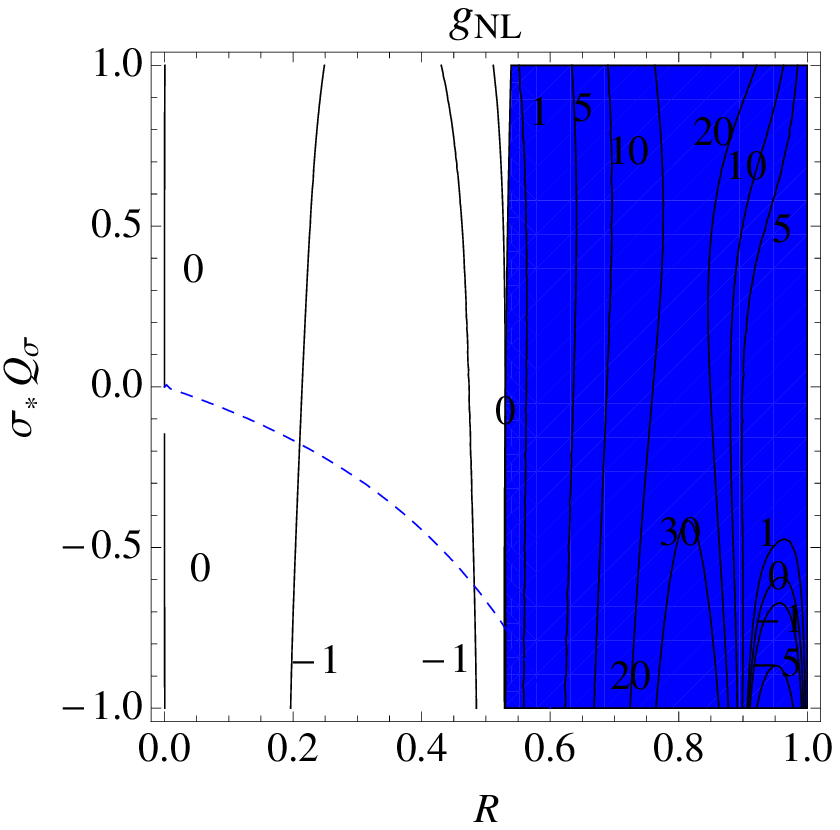}  
  \end{tabular}
  \end{center}
  \caption{The contour plots on $R- Q_{\sigma}\sigma_*$ plane, 
  of the power spectrum ${\cal P}_{\zeta}$  normalized by the value at $R=1$ (upper-left), $\fnl$ (upper-right), $\taunl$ (lower-left)
  and $\gnl$ (lower-right) for Case C, $\sigma_*=0.5\, \Mp$. We put $\eta_*=\xi_*=0$.
  The blue shaded region corresponds to too large tensor-to-scalar ratio to be consistent with the observation, $r_T<0.36$. Along the blue dashed line $\delta=0$.  }
  \label{fig:Case3}
\end{figure}

In the upper left panel of figures~\ref{fig:Case1} - \ref{fig:Case3}, 
 we show  the power spectrum in the plane of 
 $(R, Q_\sigma \sigma_*)$ plane in the upper-left window for each cases with different $\sigma_*$ ;
 $10^{-3}\, \Mp$ (Case A), $0.05\, \Mp$ (Case B), $0.5\, \Mp$ (Case C).
We have normalized the amplitude of the power spectrum by
 the value at the pure curvaton limit of $R=1$,  namely ${\cal P}_{\zeta}/{\cal P}_{\zeta,0}$.
The slow-roll parameter $\epsilon_*$ is calculated from \eq{index} considering $n_s=0.97$ and
 we used for simplicity $\eta_*=\xi_*=0$.  
We assume that the correct observational value can be attained using the residual parameter of $H_*$.
 
The three cases in figures~\ref{fig:Case1} - \ref{fig:Case3} show the following features:
\begin{itemize}
\item Case A : $\sigma_*=10^{-3}\, \Mp$ and $\tilder \gg1$ in most of the region.\\
Modulated reheating and curvaton contributions are dominant and
 inflaton contribution to the curvature perturbation is subdominant, 
\item Case B : $\sigma_*=0.05\, \Mp$ and $\tilder \sim 1$ as shown in figure~\ref{fig:tilder2}.\\
All three contributions are effective,
\item Case C : $\sigma_*=0.5\, \Mp$ and $\tilder \ll 1$  in most of the region.\\
Inflaton contribution is dominant and it scales as $(1-R)^2$.
$R<0.45$ is allowed from the constraint on the tensor-to-scalar ratio.
\end{itemize}

In the Case A (upper-left window in figure~\ref{fig:Case1}),
 $\delta\sigma$ contribution is dominant in the overall region and
 the inflaton contribution is subdominant. 
In the cancellation region, the relative contribution from $\delta \sigma$ decreases. 
Of course, by increasing $H_*$, the desired amplitude of ${\cal P}_{\zeta}$ can be recovered. 
The tensor-to-scalar ratio is always much smaller than 0.01 in the shown parameter range.

In the Case B (upper-left window in figure~\ref{fig:Case2}), 
 the inflaton contribution is effective for cancellation region and
 modulation and/or curvaton effects are effective in the rest. 
The cancellation between the modulated reheating and the curvaton appears
 around $R\sim 0.7$ for negative $Q_\sigma \sigma_*$. 
The tensor-to-scalar ratio is comparable to the observation constraint and
 a region around $(R,\Qs) \sim (0.6,-1)$ is ruled out as shown in the right window of figure~\ref{fig:tilder2}. 

In the Case C (upper-left window in figure~\ref{fig:Case3}), 
 the density perturbation dominantly comes from the inflaton field
 but the magnitude changes due to the effect of curvaton. 
 For large $R$ the magnitude decreases.
 A portion of parameter space is excluded because of too large tensor-to-scalar ratio.

\subsection{Non-Gaussianity}
When the curvaton field dominates the energy density when it decay, i.e. $R \sim 1$,
 the curvature perturbation is dominated by the pure curvaton and the non-Gaussianity is suppressed.
However in the other case $0 \lesssim R \ll1$, 
 there is a possibility to get large non-Gaussianity.

The bispectrum $B_\zeta$ is given by
\dis{
\VEV{\zeta_{\vec k_1}\zeta_{\vec k_2}\zeta_{\vec k_3} } =(2\pi)^3 B_\zeta(k_1,k_2,k_3) \delta(\vec{k_1}+\vec{k_2}+\vec{k_3}) ,
} 
 and the dimensionless non-linearity parameter for the bispectrum, $\fnl$, is defined by
 \dis{
 B_{\zeta}(k_1,k_2,k_3) =\frac65 \fnl [P_\zeta(k_1)P_\zeta(k_2)+P_\zeta(k_2)P_\zeta(k_3)+P_\zeta(k_3)P_\zeta(k_1) ]. \label{fnldefine}
 }
 
 From the trispectrum $T_\zeta$ given by
 \dis{
\VEV{\zeta_{\vec k_1}\zeta_{\vec k_2}\zeta_{\vec k_3}\zeta_{\vec k_4} }_c =(2\pi)^3 T_\zeta(k_1,k_2,k_3,k_4) \delta(\vec{k_1}+\vec{k_2}+\vec{k_3}+\vec{k_4}) ,
 }
  the dimensionless non-linearity parameters $\taunl$ and $\gnl$ are defined as~\cite{astro-ph/0611075}
 \dis{
 T_\zeta(k_1,k_2,k_3,k_4)=&\taunl [ P_\zeta(k_{13})P_\zeta(k_3)P_\zeta(k_4) 
 + 11 \text{ perms} ]\\
& + \frac{54}{25}\gnl[ P_\zeta(k_{2})P_\zeta(k_3)P_\zeta(k_4) + 3 \text{ perms}  ].
\label{taunlgnldefine}
 }

The current bounds have been derived by several groups~\cite{Smidt:2010ra,Fergusson:2010gn}.
For instance, Smidt {\it et. al.} reported as
 $-7.4 < g_{\rm NL}10^{-5} < 8.2$ and $-0.6< \tau_{\rm NL} 10^{-4} <3.3$~\cite{Smidt:2010ra}.

In figure~\ref{fig:Case1}, we show the contours of the non-linearity parameters $\fnl$ (upper-right window),
 $\taunl$ (lower-left window) and $\gnl$ (lower-right window) for the Case A.
For the contours we have taken account the relations,
 $Q_{\sigma\sigma}\sigma_*^2 = Q_{\sigma}\sigma_* (6 Q_{\sigma}\sigma_* +1) $ and
 $Q_{\sigma\sigma\sigma} \sigma_*^3=12\Qs Q_{\sigma\sigma} \sigma_*^3$, 
 which are motivated from the specific example we will show in the next section.
We can see clearly the enhancement of non-linearity parameters in the cancellation region
 of modulated reheating and curvaton. The pure curvaton limit is recovered along the line of $\Qs=0$.

In figure~\ref{fig:Case2} for the Case B,
 we can see the enhancement of the non-linearity parameters along the cancellation region. 
The difference from the Case A is that now
 the inflaton becomes more important and diminishes the non-Gaussinaity.

In both cases of A and B,
 the large non-Gaussianity is dominantly due to $\delta \sigma$, while the power spectrum comes from both depending on $\tilder$, 
as explained in the Appendix. 
 The non-linearity parameters can be enhanced around the cancellation region ($\delta \ll 1$ and $\delta \ll \tilder$). 
In this region,
 a large $\fnl$ comes dominantly from $\zeta_{2,\sigma\sigma}$ as defined in \eq{zeta123A} of the Appendix and estimated to be
\dis{
\fnl\simeq \frac56\frac{\tilder^2}{(1+\tilder)^2} \frac{1}{\delta^2} \frac{9(1-R)\Qss -6R +R(1-R)(3+R)(\Qs-2)^2 }{4R^2}.
}
Therefore small $\delta$ (large cancellation due to fine-tuning) can induce larger $\fnl$.
However note that $\tilder$ is proportional to $\delta^2$ (so $\tilder^2\propto \delta^4$), thus
 too small $\delta$ makes $\tilder$ becomes smaller than $\delta$ itself and reduces $\fnl$. 
One can see this behavior clearly in the upper right figure of figure~\ref{fig:Case2}:
$\fnl$ decreases when approaching the cancellation line (blue dashed line).

In the same region $\taunl$ is also dominated by $\zeta_{2,\sigma\sigma}$ term and approximately the squared of $\fnl$,
\dis{
\taunl \simeq \bfrac{1+\tilder}{\tilder}\left(\frac65\fnl \right)^2.
}
Large $\gnl$ is possible in the same cancellation region dominated by $\zeta_{3,\sigma\sigma\sigma}$ term and  given by
\dis{
\gnl\simeq \frac{25}{54} &\frac{\tilder^3}{(1+\tilder)^3}\frac{1}{\delta^3}  \frac{1}{8R^3} \\
 \times[&9(1-R) \Qsss +9R(1-R)(3+R)(3\Qs-2)(3\Qss +2) \\
 &+ 36R +R(1-R)(3+R) (-3+4R+3R^2)(3\Qs-2)^3  ].
}
As you can see here, $\gnl$ has different sign in the opposite side of the cancellation line due to the odd exponent '3' of $\delta$.

Here one can see that
 there are two points in both sides of $3 Q_{\sigma} \sigma_* \simeq -2R/(1-R)$ 
 those give the same values of  ${\cal P}_{\zeta}$ and $f_{\rm NL}$.
This means that the measurements of only ${\cal P}_{\zeta}$ and $f_{\rm NL}$
 can not determine $R$ and $Q_{\sigma}\sigma_*$ uniquely.
However, this degeneracy can be resolved by measuring $g_{\rm NL}$
 because one predicts $g_{\rm NL}$ to be positive and the other does it to be negative.

In figure~\ref{fig:Case3} for the Case C,
 the region of large non-linearity parameters are in the region of $R> 0.5$
 but which is excluded out by large $r_T$.

\section{A simple model}
\label{models}

As we have seen in the previous sections,
 if a light scalar curvaton field $\sigma$ has an interaction with inflaton field $\phi$,
 the fluctuation of curvaton field $\delta\sigma$ can modulate reheating through the decay of inflaton
 and can affect the curvature perturbation besides the usual curvaton mechanism. 
The resultant power spectrum and the non-linearity parameters can be considerably affected.
In this section, we examine these effects in a simple model.

The inflaton $\phi$ and the curvaton $\sigma$ can have the following interactions  in the scalar potential
\begin{equation}
 V[\phi,\sigma] = \frac{1}{2} m_{\phi}^2 \phi^2 +  \frac{1}{2} m_{\sigma}^2 \sigma^2
  + \frac{1}{2}\lambda_{\phi\sigma} \phi^2 \sigma^2,
\end{equation}
as well as an interaction with another scalar field as given in \eq{PhiSigmaInteraction}.
It is also possible that the inflaton has interaction with other fields independent from the curvaton.
We consider that the inflation is dominantly driven by a single inflaton field
 with its quadratic mass term potential
 by assuming $m_{\phi} \gg m_{\sigma}$ and 
\begin{equation}
m_{\phi}^2 \gg \lambda_{\phi\sigma}\sigma_*^2.
\label{Cond:SmallLambdaPhiSigma}
\end{equation}
We consider cases of vanishing $\lambda_{\phi\sigma}$ in subsection~\ref{sebsec:modelI} and \ref{sebsec:modelII},
 and mention the effect of nonvanishing $\lambda_{\phi\sigma}$ in subsection~\ref{sebsec:modelIII}.
Its interactions with other fields are important during reheating or later.
Therefore during inflation, field equations are reduced to
\begin{eqnarray} 
 && H^2 = \frac{1}{6 M_P^2} m_{\phi}^2 \phi^2, \\
 && 3 H \dot{\phi} + m_\phi^2\phi = 0 ,
\end{eqnarray}
 under the slow-roll condition
 \dis{
 \epsilon\equiv \frac{\Mp^2}{2} \bfrac{V_\phi}{V}^2\simeq \frac{2\Mp^2}{\phi^2} \ll 1, \qquad
 \eta \equiv \frac{M_P^2 V_{\phi\phi}}{V} \ll 1 ,
 }
  and the inflaton-domination condition
\begin{equation}
 \frac{1}{2} m_{\phi}^2 \phi^2 \gg \frac{1}{2} m_{\sigma}^2 \sigma^2 .
\end{equation}
By solving field equations, we obtain 
\begin{eqnarray} 
 \phi_*^2 =  ( 4 N_{\rm inf} +2 ) M_P^2,
\end{eqnarray}
 where $N_{\rm inf}$ is the number of e-fold at horizon exit from the end of inflation.
The power spectrum is
\begin{equation}
 {\cal P}_{\zeta} \simeq
 \frac{(1-R)^2}{6 (2 \pi)^2} \frac{m_{\phi}^2}{M_P^2}( 2 N_{\rm inf} +1 )^2 + 
 \left\{ (1-R)Q_\sigma\sigma_* + \frac{2R}{3} \right\}^2 \frac{m_{\phi}^2}{3(2\pi\sigma_*)^2}( 2 N_{\rm inf} +1 ) .
\end{equation}
The inflaton mass $m_{\phi}$ controls the amplitude of the density perturbation.
For the observed ${\cal P}_{\zeta}$ we can estimate  $m_{\phi}$.
\begin{figure}[!t]
  \begin{center}
  \begin{tabular}{c c}
   \includegraphics[width=0.5\textwidth]{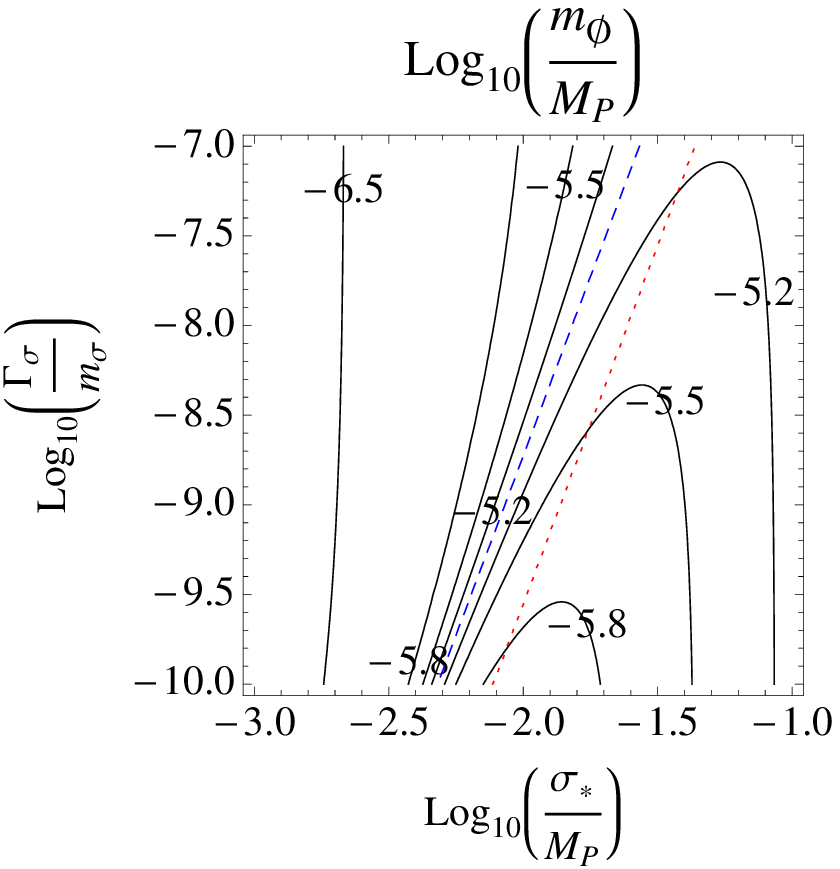}
    & 
  \includegraphics[width=0.5\textwidth]{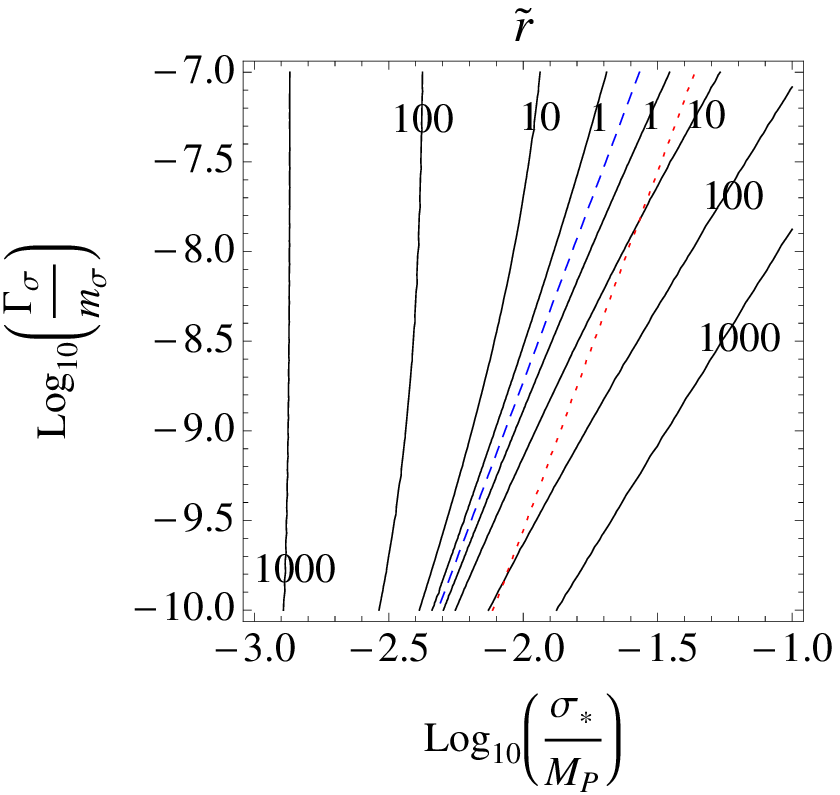}
  \end{tabular}
  \end{center}
  \caption{{\bf [Left window]} : The contour plot of $ m_{\phi} $ of the Model I
 in the unit of $\log_{10} \left(\frac{m_{\phi}}{M_P}\right)$ 
 where ${\cal P}_{\zeta}$ has been fixed by the observed value ${\cal P}_{\zeta}=2.44\times10^{-9}$.
 In the left side of the red line radiation is dominated at the time of curvaton decay and in the right side curvaton dominates. 
 The cancellation between curvaton and modulated reheating occurs along the blue line where $\tilder=0$. 
 {\bf [Right window]} : The contour plot of $\tilder$. The red and blue lines are the same as in the left window.}
  \label{fig:InflatonMassA}
\end{figure}

It is natural to assume that inflaton decays at its oscillating stage after a while,
 so that  $\Gamma_{\phi}(\sigma)/m_{\phi}$ is very small. 
In this case $Q$ is well approximated by~\cite{Suyama:2007bg}
\dis{
Q\simeq -\frac16 \log\bfrac{\Gamma_{\phi}(\sigma)}{H_c}.
\label{Qdef}
}
The derivatives are expressed as
\begin{eqnarray}
Q_\sigma
 &=& -\frac16\frac{\partial_\sigma \Gamma_\phi}{\Gamma_\phi}, \\ 
Q_{\sigma\sigma} &=&-\frac16 \left( \frac{\partial_\sigma^2 \Gamma_\phi}{\Gamma_\phi}
 -\frac{ (\partial_\sigma \Gamma_\phi)^2}{\Gamma^2_\phi} \right), \\
Q_{\sigma\sigma\sigma} &=&-\frac16 \left( \frac{\partial_\sigma^3 \Gamma_\phi}{\Gamma_\phi}
 -3\frac{ \partial_\sigma \Gamma_\phi \partial^2_\sigma \Gamma_\phi  }{\Gamma^2_\phi} 
+ 2\frac{ (\partial_\sigma \Gamma_\phi)^3 }{\Gamma^3_\phi}  \right) .
\label{Qderivatives}
\end{eqnarray}
%

\subsection{
Model I :  Inflaton decays through only the coupling in \eq{PhiSigmaInteraction} }
\label{sebsec:modelI}

\begin{figure}[!t]
  \begin{center}
  \begin{tabular}{c}
   \includegraphics[width=0.5\textwidth]{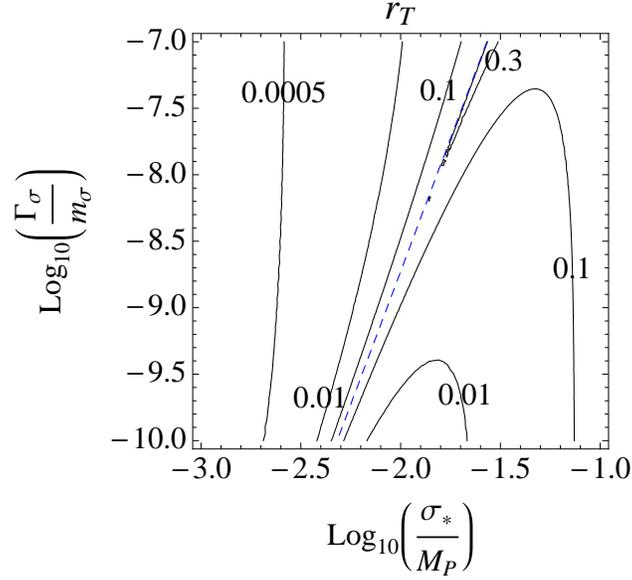}
  \end{tabular}
  \end{center}
  \caption{The contour plot of $r_T$ for the Model I. The blue line shows $\tilder=0$.  }
  \label{fig:Model1_rT}
\end{figure}

First, let us consider the case that the inflaton $\phi$ decays through only 
 the coupling in \eq{PhiSigmaInteraction}.
For this case, we obtain
\begin{equation}
 (Q_{\sigma}\sigma_*, Q_{\sigma\sigma}\sigma_*^2, Q_{\sigma\sigma\sigma}\sigma_*^3 )
 = \left(
\begin{array}{ccc}
 -\frac{1}{3} , & \frac{1}{3} , & -\frac{2}{3}
\end{array}
 \right) .
\end{equation}
 The parameter defined in \eq{R} at the time of curvaton decay is can be obtained 
 using \eqs{rhor}{rhosigma} by
\begin{eqnarray}
 R = 
\frac{ \sigma_*^2/\Mp^2 }{8 (a_{m_{\sigma}}/a_{\Gamma_{\sigma}})+ \sigma_*^2/\Mp^2},
\end{eqnarray}
 with 
\begin{equation}
 \frac{a_{m_{\sigma}}}{a_{\Gamma_{\sigma}}} = 
 \left\{
\begin{array}{l}
   \left( \frac{\Gamma_{\sigma}}{m_{\sigma}} \right)^{1/2}
   \qquad\qquad\qquad {\rm radiation-dominated}  \\
   \left( \frac{ \sigma_*^2 } {6 M_P^2}\right)^{-1/3}
  \left( \frac{\Gamma_{\sigma}}{m_{\sigma}} \right)^{2/3} \quad\quad \sigma{\rm-dominated }
\end{array}
\right. . \label{scale_A}
\end{equation}
Here $\Gamma_\sigma$ is the decay rate of the curvaton field and
 $a_{m_{\sigma}}/a_{\Gamma_{\sigma}}$  is expressed in different ways depending on
 whether it is radiation-dominated or curvaton-dominated when the curvaton decays at $H=\Gamma_\sigma$.

In this case, there are only three parameters $m_{\phi}, \sigma_*$ and $\Gamma_{\sigma}/m_{\sigma}$, 
 since $Q_{\sigma}\sigma_*$ and so on are completely fixed.
As mentioned above, from the normalization of the power spectrum, 
 $m_{\phi}$ can be expressed by the other two parameters
 as shown in the left window of figure~\ref{fig:InflatonMassA}. 
The largest value to $m_{\phi}$ in figure~\ref{fig:InflatonMassA}  seems to be $10^{-5.2} M_P$
 which is the same as the results in the quadratic chaotic inflation.
This is because the region corresponds to the cancellation, $3 Q_{\sigma} \sigma_* \simeq -2R/(1-R)$ line (blue line and $R=1/3$ in this case),
where $\tilder \simeq 0$ and the contribution  from $\sigma$ field cancels, and the dominant contribution comes from $\phi$ field.
On the other hand, $\sigma$ contribution is not negligible in other regions and hence
 a smaller $m_{\phi}$ is needed to produce the observable ${\cal P}_{\zeta}$.
In the right window of figure~\ref{fig:InflatonMassA}, we showed a contour plot of $\tilder$.
We note that
 the condition for a negligible $\lambda_{\phi\sigma}$, Eq.~(\ref{Cond:SmallLambdaPhiSigma}),
 can be rewritten as
\begin{equation} 
\lambda_{\phi\sigma} \ll 10^{-6}\left(\frac{m_{\phi}}{10^{-5} M_P}\right)^2 \left( \frac{10^{-2} M_P}{\sigma_*} \right)^2 .
\end{equation}

In figure~\ref{fig:Model1_rT}, we showed the contour plot of $r_T$. The present bound on the $r_T$ does not constrain this model, however future bound can rule out the parameter range around the cancellation region along the blue line.

\begin{figure}[t]
\begin{center}
\begin{tabular}{cc}
 \begin{minipage}{0.5\hsize}
  \begin{center}
    \centerline{\includegraphics[width=80mm]{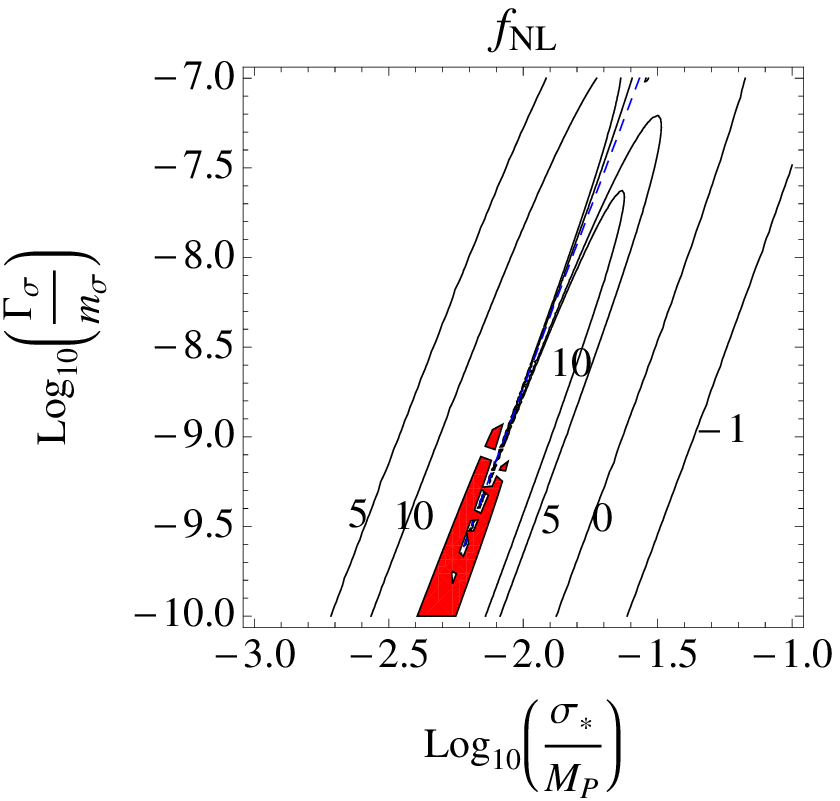}}
  \end{center}
 \end{minipage}
 \begin{minipage}{0.5\hsize}
  \begin{center}
     \centerline{\includegraphics[width=80mm]{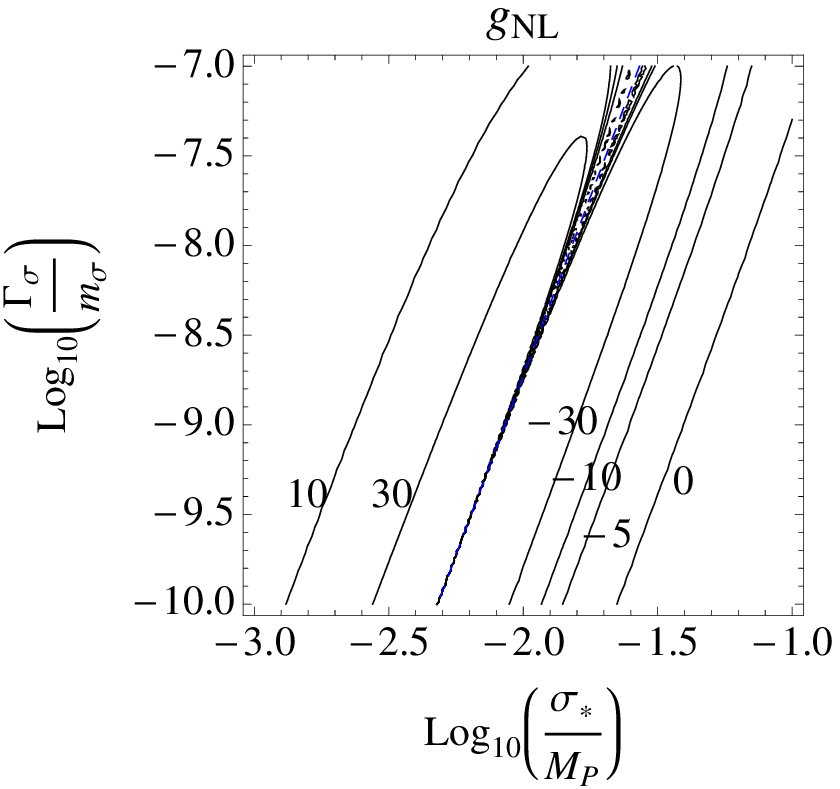}}
  \end{center} 
 \end{minipage}
\end{tabular}
 \caption{{\bf [Left window]} : The contour plot 
 of $ f_{\rm NL} $  in $(\sigma_*/\Mp, \Gamma_{\sigma}/m_{\sigma} )$ plain
 for the Model I.
 The red shaded region corresponds to too large $f_{\rm NL}$ to be consistent with the observation.
 {\bf [Right window]} : The contour of $g_{\rm NL}$ for the Model I.
}
\label{fig:A:fNLandgNL}
\end{center}
\end{figure}

In figure~\ref{fig:A:fNLandgNL}, the non-linearity parameters $\fnl$ (left) and $\gnl$ (right) are shown.

%

\subsection{ Model II : Inflaton decays through the coupling in \eq{PhiSigmaInteraction} and others }
\label{sebsec:modelII}

\begin{figure}[t]
\begin{center}
\begin{tabular}{cc}
 \begin{minipage}{0.5\hsize}
  \begin{center}
    \centerline{\includegraphics[width=80mm]{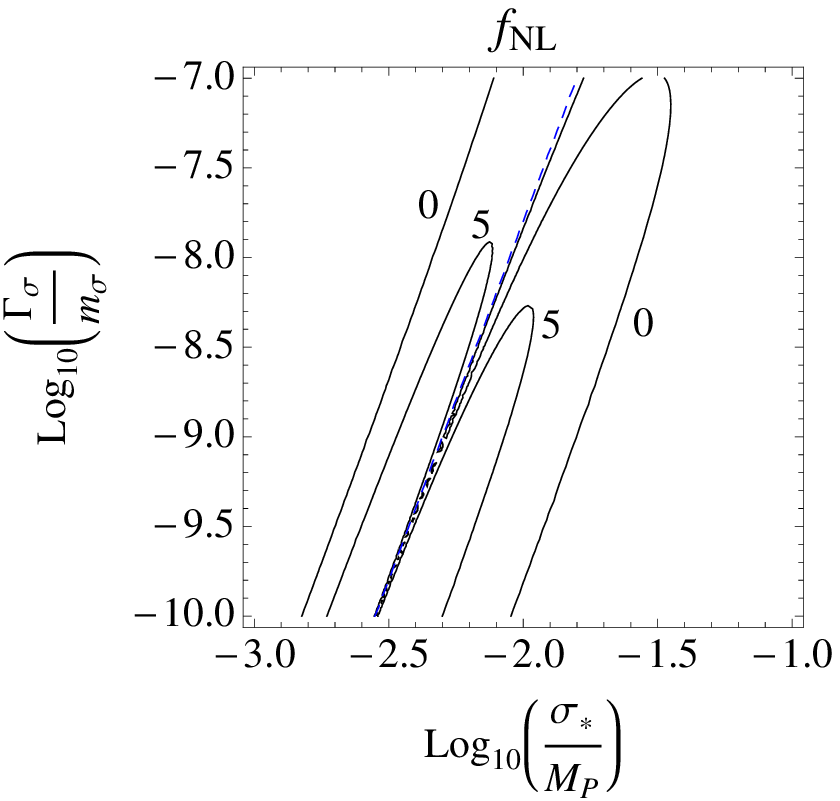}}
  \end{center}
 \end{minipage}
 \begin{minipage}{0.5\hsize}
  \begin{center}
     \centerline{\includegraphics[width=80mm]{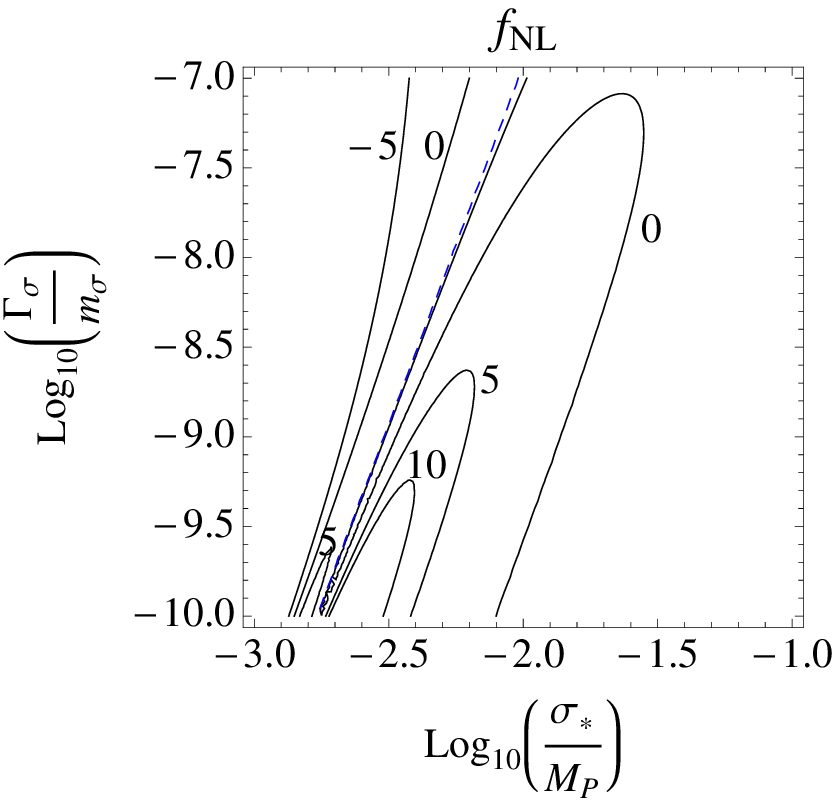}}
  \end{center} 
 \end{minipage}
\end{tabular}
\caption{{\bf [Left window]} : The contour plot of $f_{\rm NL}$ with $Br=0.3$
 on $(\sigma_*/\Mp,  \Gamma_{\sigma}/m_{\sigma} )$ plane for the Model II.
 The red shaded region corresponds to too large $f_{\rm NL}$ to be consistent with the observation.
 {\bf [Right window]} :  The same as left window but with $Br=0.1$.
 }
\label{fig:B:fN}
\end{center}
\end{figure}

Next, let us consider the case that the inflaton $\phi$ has
 a nonvanishing $\sigma$ independent decay modes  and
 the total decay width is given by Eq.~(\ref{Deay:total}).
For this case, we obtain
\begin{equation}
 (Q_{\sigma}\sigma_*, Q_{\sigma\sigma}\sigma_*^2, Q_{\sigma\sigma\sigma}\sigma_*^3 )
 = \left(
\begin{array}{ccc}
 -\frac{1}{3}Br , & \frac{1}{3}Br (2 Br - 1) , & -\frac{2}{3} Br^2 ( 4 Br -3 )
\end{array}
 \right) ,
\end{equation}
 with $Br \equiv \Gamma_{\phi}^{(CD)}/\Gamma_{\phi} $.
Obviously, the $Br \rightarrow 1$ limit reduces to the Model I in the previous subsection. 
In this case, we have four parameters $m_{\phi}, \sigma_*, \Gamma_{\sigma}/m_{\sigma}$ and $Br$. 
Again,  $m_{\phi}$ can be used for the normalization of the power spectrum. 
In figure~\ref{fig:B:fN}, we show the contour plot of $\fnl$ for $Br=0.3$ (left) and $0.1$ (right).
As we have seen, the cancellation happens at $3 Q_{\sigma} \sigma_* \simeq -2R/(1-R)$
or  $R\simeq Br/(2+Br)$.

\subsection{Model III :  The effective inflaton mass with the coupling in \eq{NRInteraction} }
\label{sebsec:modelIII}

Finally, let us consider the case of nonvanishing $\lambda_{\phi\sigma}$. 
Then, through just only the coupling (\ref{NRInteraction}),
 in other words even without direct coupling (\ref{PhiSigmaInteraction}),
 the inflaton decay width
 $\Gamma \sim M_{\phi}^3/M_P^2$
 with the effective inflaton mass $M_{\phi}^2 = m_{\phi}^2 + \lambda_{\phi\sigma}\sigma_*^2$
 which  is $\sigma$ dependent.

For this case, we obtain
\begin{equation}
 (Q_{\sigma}\sigma_*, Q_{\sigma\sigma}\sigma_*^2, Q_{\sigma\sigma\sigma}\sigma_*^3 )
 = \left(
\begin{array}{ccc}
 -\frac{1}{2} Fr , & \frac{1}{2} Fr (2 Fr - 1) , &  Fr^2 (3 - 4Fr )
\end{array}
 \right) ,
\end{equation}
 with $Fr \equiv \lambda_{\phi\sigma}\sigma_*^2/M_{\phi}^2 $.
The qualitative behavior of this model is the same as that of the Model II up to numerical factors,
 by replacing $Br$ with $Fr$.

\section{Conclusion}
\label{conclusion}
 
We have studied the case where a light scalar field $\sigma$
 induces the modulated reheating by the inflaton decay
 and also acts as the curvaton by its late time decay in the presence of the curvature perturbation generated from the inflaton field itself.
In fact, the coupling in \eq{PhiSigmaInteraction} is possible from the gauge invariance of
 the SM~\footnote{Of course, it is also possible that those terms are forbidden
 by additional symmetry such as a certain $Z_2$-parity.},
 provided both $\phi$ and $\sigma$ are gauge singlet
 as is often assumed to preserve the flatness of the potential.
 
When $\sigma$ field contributes to both
 inducing the modulated reheating by the inflaton decay
 and generating the density perturbation as the curvaton,
 there could be a cancellation between two contributions
 along the line $3 Q_{\sigma} \sigma_* \simeq -2R/(1-R)$.
Around this cancellation region,
 $\delta\sigma$ contribution to the power spectrum of the density perturbation is subdominant
 and the inflaton contribution becomes dominant.
Near such a parameter region, mostly the middle range of $R$ and a negative $Q_{\sigma}\sigma_*$,
 non-linearity parameters tend to be large
 because of the cancellation between the modulated reheating and the curvaton
 originated from the same field $\sigma$.
In this sense, this cancellation is a kind of mechanisms to generate a large non-Gaussianity

As specific models, we have also studied a quadratic inflation and curvaton model
 and demonstrated
 how the parameter space of a given inflaton and curvaton model would be constrained
 by taking the interaction between them into account.
The measurement of non-linearity and tensor-to-scalar ratio 
 may probe the strength of (non-)interaction between the inflaton and the curvaton.

\section*{Acknowledgments}

K.-Y.C is supported by the 
National Research Foundation of Korea (NRF) grant funded by the 
Korea government (MEST) (No. 2011-0011083).
K.-Y.C acknowledges the Max Planck Society (MPG), the Korea Ministry of
Education, Science and Technology (MEST), Gyeongsangbuk-Do and Pohang
City for the support of the Independent Junior Research Group at the Asia Pacific
Center for Theoretical Physics (APCTP).
This work of O.S. is in part supported by
 the scientific research grants from Hokkai-Gakuen. 
O.S. would like to thank the APCTP for warm hospitality during his stay
 where this work has been completed.

\appendix
\section{Large non-Gaussianity}
The curvature perturbation $\zeta$ can be decomposed of the contributions from each field perturbations order by order  as
\begin{equation}
 \zeta = \zeta_1+\frac{1}{2}\zeta_2+\frac{1}{6}\zeta_3+\ldots,
\label{zeta_A}
\end{equation} 
with
\dis{
\zeta_1=& \zeta_{1,\phi}\delta\phi + \zeta_{1,\sigma}\delta\sigma,\\
\zeta_2=& \zeta_{2,\phi\phi}(\delta\phi)^2 +2 \zeta_{2,\phi\sigma}(\delta\phi)(\delta\sigma)+ \zeta_{2,\sigma\sigma}(\delta\sigma)^2,\\
\zeta_3=& \zeta_{3,\phi\phi\phi}(\delta\phi)^2 + 3\zeta_{3,\phi\phi\sigma}(\delta\phi)^2(\delta\sigma) + 3\zeta_{3,\phi\sigma\sigma}(\delta\phi)(\delta\sigma)^2 + \zeta_{3,\sigma}(\delta\sigma)^3.\\
\ldots& \label{zeta123A}
}
Here we considered two-field case for simplicity and we suppressed $*$ which denotes the value at horizon exit,
 i.e. $\delta\phi=\delta\phi_*$ and $\delta\sigma=\delta\sigma_*$. For the model of modulated reheating by curvaton, we can read each component from \eq{zeta3}.
 
The power spectrum is given from \eq{DefPzeta} by
\dis{
{\cal P}_\zeta= (\zeta_{1,\phi}^2 + \zeta_{1,\sigma}^2) \bfrac{H_*}{2\pi}^2 = \zeta_{1,\phi}^2(1 + \tilder) \bfrac{H_*}{2\pi}^2, 
}
where we used \eq{Pexit} and $\tilder\equiv \zeta_{1,\sigma}^2 / \zeta_{1,\phi}^2$.
From the definitions in \eq{fnldefine} and \eq{taunlgnldefine}, the non-Gaussianity parameters are given by
\dis{
\fnl =& \frac{5}{6} \frac{\zeta_{1,\phi}^2 \zeta_{2,\phi\phi} +  2\zeta_{1,\phi}\zeta_{1,\sigma}\zeta_{2,\phi\sigma}  +\zeta_{1,\sigma}^2\zeta_{2,\sigma\sigma}  }{(\zeta_{1,\phi}^2 + \zeta_{1,\sigma}^2)^2},\\
\taunl=& \frac{ \sum_{a,b,c} \zeta_{1,b}\zeta_{1,c} \zeta_{2,ab}\zeta_{2,ac} }{(\zeta_{1,\phi}^2 + \zeta_{1,\sigma}^2)^3}, \\
\gnl=&\frac{25}{54} \frac{ \sum_{a,b,c} \zeta_{1,a}\zeta_{1,b}\zeta_{1,c} \zeta_{3,abc} }{(\zeta_{1,\phi}^2 + \zeta_{1,\sigma}^2)^3}.\label{nonG}
}
Here $a,b,c$ denotes $\phi$ and $\sigma$.

First we will investigate the condition for large $\fnl$ in the curvaton-modulated scenario.
Using \eq{zeta3}, the denominator of $\fnl$ in \eq{nonG} becomes
\dis{
(\zeta_{1,\phi}^2 + \zeta_{1,\sigma}^2)^2 = \left[ \frac{1}{2\Mp^2\epsilon_*}(1-R)^2 (1+\tilder) \right]^2 \sim \frac{(1+\tilder)^2}{\Mp^4\epsilon_*^2},
}
where in the last equation we dropped $\mathcal{O}(1)$ coefficient, $R$ and $Q_ \sigma \sigma $. 
In the same way, the numerators are
\dis{
\zeta_{1,\phi}^2 \zeta_{2,\phi\phi}  =& \frac{1}{4\Mp^4\epsilon_*^2}(1-R)^3[R(R+3)+2\epsilon_* - \eta_*] \sim \frac{1}{\Mp^4\epsilon_*^2},\\
2\zeta_{1,\phi}\zeta_{1,\sigma}\zeta_{2,\phi\sigma}  =& \frac{1}{9\Mp^2\epsilon_*\sigma^2} R(1-R)^2(3+R)[3Q_\sigma \sigma (1-R)+2R][3Q\sigma - 2] 
\sim \frac{1}{\Mp^4\epsilon_*^2} \frac{\tilder}{\delta},\\
\zeta_{1,\sigma}^2\zeta_{2,\sigma\sigma} =& \frac{1}{81\sigma^4}[3Q_\sigma \sigma (1-R)+2R]^2[9Q_{\sigma\sigma} \sigma^2 (1-R) -6R +R(1-R)(3+R) (3Q_\sigma\sigma-2)^2]\\
\sim& \frac{1}{\Mp^4\epsilon_*^2} \frac{\tilder^2}{\delta^2}.
}
Therefore we find that large $\fnl$ can arise from the last term $\zeta_{1,\sigma}^2\zeta_{2,\sigma\sigma}$ to give
\dis{
\fnl \sim  \frac{\tilder^2}{(1+\tilder)^2}\frac{1}{\delta^2}.
}
It is easy to see that
\dis{
\fnl \sim \frac{1}{\delta^2}\qquad \text{for}\qquad \tilder \gtrsim 1,  \qquad \text{and} \qquad \delta \ll1.
}
and 
\dis{
\fnl \sim \frac{\tilder^2}{\delta^2}\qquad \text{for}\qquad  \tilder \lesssim 1,  \qquad \text{and} \qquad \delta \ll \tilder.
}
However $\tilder $  and $\delta$ are not independent variables and $\tilder$ is proportional to $\delta^2$ in our case as in \eq{tilderdelta}.  
Therefore for a given parameters, $\fnl$ vanishes in the  limit of  $\delta \rightarrow 0$ when $\tilder$ becomes vanishing too.
The large $\fnl$ can arise whenl $\delta$ is smaller than 1 and smaller than $\tilder$.

Similarly, large $\taunl$ and $\gnl$ are obtained in the same region as that of $\fnl$
 when $\zeta_{2,\sigma\sigma}$ and $\zeta_{3,\sigma\sigma\sigma}$ term dominates and
 are estimated to be
\dis{
\taunl \sim \frac{\tilder^3}{(1+\tilder)^3} \frac{1}{\delta^4},\qquad \text{and}\qquad  \gnl\sim \frac{\tilder^3}{(1+\tilder)^3} \frac{1}{\delta^3}. 
} 
We find that $\gnl$ is smaller than $\taunl$ in the large non-Gaussianity region.



\end{document}